\documentclass[%
 reprint,
superscriptaddress,
 amsmath,amssymb,
 aps,
]{revtex4-2}

\usepackage{dsfont}
\usepackage{svg}
\usepackage{graphicx}
\usepackage{dcolumn}
\usepackage{physics}
\usepackage{amsmath}
\usepackage{bbold}
\usepackage{float}

\usepackage{hyperref}
\hypersetup{
    colorlinks=true,
	citecolor=blue,
    linkcolor=red,
	urlcolor = blue
}

\usepackage{bm}
\usepackage[mathlines]{lineno}%

\newcommand{\tauL}{\tau^{\mathcal{L}}_{\rm m}}

\begin{document}
\title{Non-Markovian Quantum Mpemba effect}%
\author{David J. Strachan}
\email{david.strachan@bristol.ac.uk}
\affiliation{\textit{H. H. Wills Physics Laboratory, University of Bristol, Bristol BS8 1TL, United Kingdom}}
\author{Archak Purkayastha}
\email{archak.p@phy.iith.ac.in}
\affiliation{\textrm{Department of Physics, Indian Institute of Technology, Hyderabad 502284, India}}
\author{Stephen R. Clark}
\email{stephen.clark@bristol.ac.uk}
\affiliation{\textit{H. H. Wills Physics Laboratory, University of Bristol, Bristol BS8 1TL, United Kingdom}}

\date{\today}   
\begin{abstract}
    Since its rediscovery in the twentieth century, the Mpemba effect, where a far-from-equilibrium state may relax faster than a state closer to equilibrium, has been extensively studied in classical systems and has recently received significant attention in quantum systems. Many theories explaining this counter-intuitive behavior in classical systems rely on memory effects. However, in quantum systems, the relation between the Mpemba effect and memory has remained unexplored. In this work, we consider general non-Markovian open quantum systems and reveal new classes of quantum Mpemba effects, with no analog in Markovian quantum dynamics. Generically, open quantum dynamics possess a finite memory time and a unique steady state. Due to non-Markovian dynamics, even if the system is initialized in the steady state it can take a long time to relax back. We find other initial states that reach the steady state much faster. Most notably, we demonstrate that there can be an initial state in which the system reaches the steady state within the finite memory time itself, therefore giving the fastest possible relaxation to stationarity. We verify the effect for quantum dot systems coupled to electronic reservoirs in equilibrium and non-equilibrium setups at weak, intermediate and strong coupling, and both with and without interactions. Our work provides new insights into the rich physics underlying accelerated relaxation in quantum systems.
\end{abstract}
\maketitle

In 1963, high school student Erasto B. Mpemba \cite{EBMpemba_1969} rediscovered an intriguing phenomenon while making icecream, previously observed by Aristotle \cite{10.1093/acprof:oso/9780199206704.001.0001} and later discussed by Descartes \cite{Descartes1950-DESDOM}, where a hot liquid mixture freezes faster than an identical cold mixture. The term {\em Mpemba} effect (MpE) has since been coined to describe the phenomenon that a far-from-equilibrium state can relax to equilibrium faster than a state closer to equilibrium. It has been studied extensively in many classical systems \cite{article4, baity2019mpemba,PhysRevE.102.012906,PhysRevLett.119.148001,article,Biswas2023MpembaEI,Busiello_2021}. Different theories exist in various contexts explaining this behavior, suggesting that it is likely not one effect but rather a broad umbrella for many mechanisms of anomalous relaxation. In classical systems, often this behavior is attributed to memory effects in dynamics, although it has been shown to exist in Markovian classical systems \cite{article2} also. Recently, numerous theoretical and experimental works have generalized MpEs to quantum systems \cite{PhysRevResearch.3.043108,Nava_2019,PhysRevLett.131.080402,PhysRevLett.127.060401,PhysRevE.108.014130,chatterjee2023multiple,zhang2024observation,wang2024mpemba,
shapira2024mpemba,joshi2024observing,rylands2023microscopic,
bao2022accelerating,Ares_2023,Kochsiek_2022,caceffo2024entangled}, considering either Markovian open quantum dynamics or isolated systems at zero temperature. However, the interplay of memory and MpE in quantum systems has remained unexplored. In this work, we investigate the possibility of fast relaxation of specific initial states in general non-Markovian open quantum systems, which have a finite, non-negligible memory time and a unique steady state. This reveals new classes of quantum MpE, with no analog in Markovian quantum dynamics.

 \begin{figure}[t]  
    \includegraphics[width=0.4\textwidth]{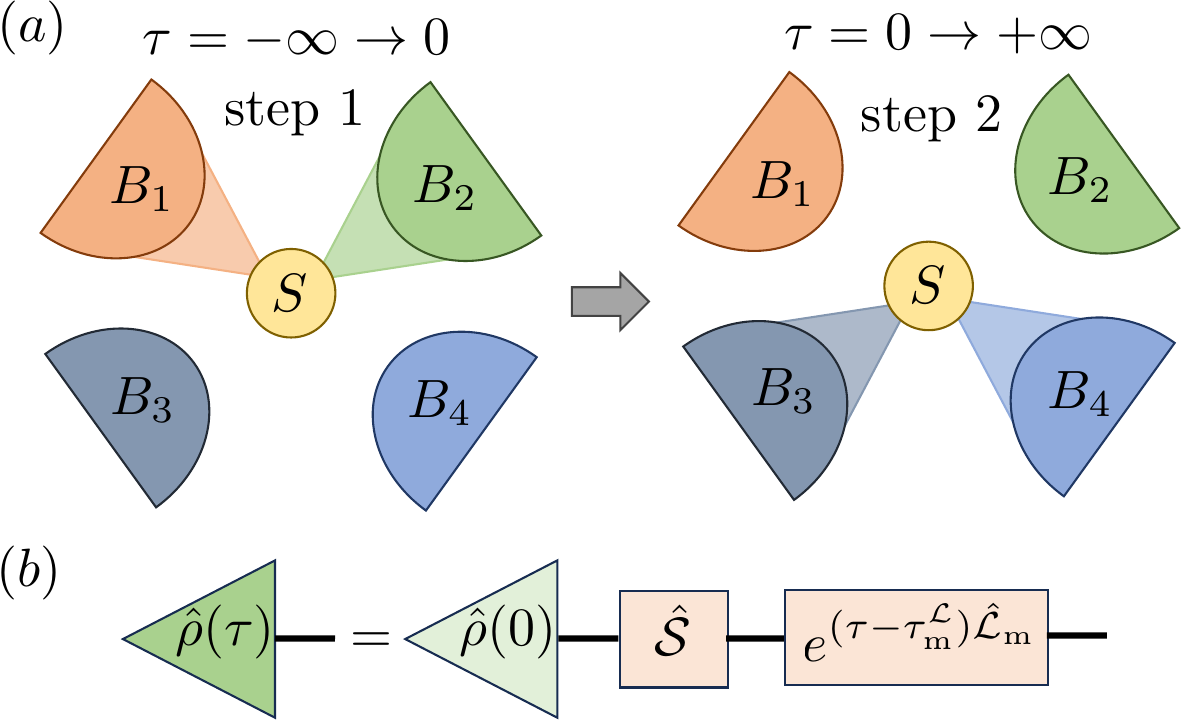}
    \caption{Setup and decomposition. (a) A system $S$ is prepared in an initial state $\hat{\rho}(0)$ that is the long-time steady state generated by coupling only to baths $B_1,B_2$ for all times $\tau < 0$. At $\tau =0$ the coupling switches to baths $B_3,B_4$. (b) The subsequent time evolution for $\tau\geq 0$ of the system density operator $\hat{\rho}(\tau)$ decomposes for $\tau > \tau^{\mathcal{L}}_{\rm m}$ into the slippage $\hat{\mathcal{S}}$ and time-independent propagator $\hat{\mathcal{L}}_{\rm m}$.}  \label{fig1}
\end{figure}

The scenario we consider is depicted in Fig.~\ref{fig1}(a). A system $S$ is coupled for times $-\infty < \tau <0$ to a set of baths such that it is prepared in the corresponding long-time steady state by $\tau=0$. For $\tau > 0$ this initial state $\hat{\rho}(0)$ is subject to time evolution generated by switching the coupling to a second distinct set of baths. Thus, the steady state of step 1 is the
initial state of step 2. If the system is finite dimensional, any given state of the system can be generated at $\tau=0$ \cite{Nielsen_and_Chuang}.  The ensuing dynamics establishing stationarity in step 2 is the focus of our work. Generically this evolution will be non-Markovian meaning that even if the first and second sets of baths are identical, so $\hat{\rho}(0)$ is already initialized to the steady state of the second baths $\hat{\rho}(\infty)$, it will be quickly perturbed away from stationarity and can take a long time to relax back. We find other special initial states $\hat{\rho}(0)$, whose preparation correspond to being stationary with first baths different from the second, which relax to the steady state of the second set of baths much faster. We call this the {\em non-Markovian quantum Mpemba effect} (NMQMpE). Our analysis is model-independent, with only a few physically motivated assumptions. We verify our findings, via numerically calculations for single and double quantum dots coupled to electronic reservoirs for equilibrium and non-equilibrium settings, at weak, intermediate and strong bath couplings, and both with and without interactions in the system.

{\it Non-Markovian dynamics---}
At $\tau=0$, there is no correlation between the system and the baths in step 2. So, we have $\hat{\rho}_{SB}(0) = \hat{\rho}(0)\hat{\rho}_{B}(0)$, where $\hat{\rho}(0)$ is the initial state of the system and $\hat{\rho}_{B}(0)$ is the combined initial state of the baths in step 2. For $\tau>0$, the dynamics of the system is given by the the completely positive trace preserving (CPTP) map, $\hat{\rho}(\tau)=\hat{\Lambda}(\tau)[\hat{\rho}(0)]$ where,
$
\hat{\Lambda}(\tau)[\bullet] = \textrm{Tr}_{B}\big(e^{-i\hat{H}\tau}\bullet\hat{\rho}_{B}(0)e^{i\hat{H}\tau}\big),
$ with $\hat{H} = \hat{H}_S+\hat{H}_{B}+\hat{H}_{SB}$, where $\hat{H}_S$ is the system Hamiltonian, $\hat{H}_{B}$ is the total Hamiltonian of the baths in step 2, and $\hat{H}_{SB}$ describes the combined system-bath coupling (setting $k_{B}=1=\hbar$). A related exact description of open quantum system dynamics is given by the time convolutionless master equation \cite{nestmann2019timeconvolutionless,chruscinski2010non,
chruscinski2022dynamical}
$\frac{\partial \hat{\rho}(\tau)}{\partial \tau}  = \hat{\mathcal{L}}(\tau)[\hat{\rho}(\tau)]$ with
$\hat{\mathcal{L}}(\tau) = \frac{d}{d\tau}[\hat{\Lambda}(\tau)]\hat{\Lambda}^{-1}(\tau)
$
from which $\hat{\Lambda}(\tau) = \mathcal{T}e^{\int_{0}^{\tau}\hat{\mathcal{L}}(\tau')d\tau'}$, where $\mathcal{T}$ is the time-ordering operator. We consider the situation where the system approaches a unique steady state in the long-time limit, i.e, $\lim_{\tau\to\infty}\hat{\Lambda}(\tau)[\hat{\rho}(0)]=\hat{\rho}(\infty)$ for any initial state $\hat{\rho}(0)$.

{\it Two different memory times---}
 A crucial difference between the above non-Markovian description and Markovian quantum dynamics is that, both $\hat{\Lambda}(\tau)$ and $\hat{\mathcal{L}}(\tau)$ are time-dependent, despite the global Hamiltonian $\hat{H}$ being time-independent. The instantaneous steady state, or the time-dependent fixed point, can be defined either as the eigenoperator of $\hat{\Lambda}(\tau)$ with eigenvalue $1$, or as the eigenoperator of $\hat{\mathcal{L}}(\tau)$ with eigenvalue $0$. Neither time-dependent fixed point may correspond to $\hat{\rho}(\infty)$ at short times, but both approach $\hat{\rho}(\infty)$ at long times. This guarantees the existence of two memory time scales $\tau^{\Lambda}_{\rm m}$, $\tauL$,
\begin{align}
& ||\hat{\Lambda}(\tau)[\hat{\rho}(\infty)] - \hat{\rho}(\infty)|| <\epsilon,~~\forall~~\tau \geq\tau^{\Lambda}_{\rm m}, \label{tau_lambda} \\
& ||\hat{\mathcal{L}}(\tau)[\hat{\rho}(\infty)]||<\epsilon~~ \forall~~ \tau \geq\tauL, \label{tau_L}
\end{align}
where $\epsilon$ is some arbitrarily small tolerance, and $|| \hat{A}||$ gives the norm of $\hat{A}$. Equation~\eqref{tau_lambda} shows that, due to non-Markovianity, even when the dynamics is initialized with $\hat{\rho}(\infty)$, it takes a time $\tau^{\Lambda}_{\rm m}$ to relax back. Equation~\eqref{tau_L} defines the time scale $\tauL$ in which the time-dependent fixed point of the propagator effectively converges to $\hat{\rho}(\infty)$. These two time scales can be different in general, which then leads to the NMQMpE as we discuss below.

{\it Unveiling the NMQMpE---}
 Although not strictly required~\footnote{The NMQME doesn't rely on the convergence of $\hat{\mathcal{L}}(\tau)$, it only requires the convergence of its fixed point $\hat{\rho}_{\textrm{TDFP}}(\tau)\to\hat{\rho}(\infty)$ where $\hat{\mathcal{L}}(\tau)[\hat{\rho}_{\textrm{TDFP}}(\tau)] = 0$.}, it is possible to argue on general grounds that  $\hat{\mathcal{L}}(\tau)$ itself converges up to an error $O(\epsilon)$ on the timescale $\tauL$, i.e, $\hat{\mathcal{L}}(\tau)=\hat{\mathcal{L}}_{\rm m} + O(\epsilon),~~\forall~~\tau \geq \tauL$, where $\hat{\mathcal{L}}_{\rm m}$ is the converged propagator \cite{Purkayastha_2021}. This gives
\begin{equation} \label{eq:3}
\hat{\Lambda}(\tau) \approx e^{(\tau-\tauL)\hat{\mathcal{L}}_{\rm m}}\hat{\mathcal{S}} \quad \forall~~\tau\geq\tauL,
\end{equation}
where $\hat{\mathcal{S}} = \hat{\Lambda}(\tauL)$, as depicted in Fig.~\ref{fig1}(b). This decomposition represents a phenomenon called initial slippage, which has been investigated in a wide class of systems, usually with weak system-bath coupling \cite{Bruch_2021,GEIGENMULLER198341,PhysRevA.28.3606,PhysRevA.32.2462,Gaspard1999SlippageOI}. It is closely associated with the assumption of a finite memory time for the environment correlations \cite{PhysRevX.11.021041,Bruch_2021}. From this decomposition, its clear $\tauL\leq\tau^{\Lambda}_{\rm m}$. Once the memory time $\tau^{\mathcal L}_{\rm m}$ is reached the dynamics can be understood in terms of the spectral decomposition of $\hat{\mathcal{L}}_{\rm m}$,
$
\hat{\mathcal{L}}_{\rm m}[\hat{\rho}]~=~\sum_{\mu}\lambda_{\mu}\hat{F}_{\mu}\textrm{Tr}(\hat{G}^{\dag}_{\mu}\hat{\rho}),
$
where $\hat{G}_{\mu}$, $\hat{F}_{\mu}$ define the damping basis for $\hat{\mathcal{L}}_{\rm m}$ which satisfy $\hat{\mathcal{L}}_{\rm m}\hat{F}_{\mu} = \lambda_{\mu}\hat{F}_{\mu}$, $\hat{\mathcal{L}}^{\dag}_{\rm m}\hat{G}_{\mu} = \lambda^{*}_{\mu}\hat{G}_{\mu}$ and the normalisation condition $\textrm{Tr}(\hat{G}_{\mu}\hat{F}_{\nu}) = \delta_{\mu\nu}$. Due to the complete positivity of $\hat{\Lambda}(\tau)$, $\textrm{Re}({\lambda_{\mu}})\leq 0$ with $\lambda_{1}=0$, $\textrm{Tr}(\hat{F}_{\mu}) = 0$ for $\lambda_{\mu} \neq 0$ \cite{Minganti_2019,10.1093/acprof:oso/9780199213900.001.0001,Rivas_2012}, where the only eigenoperator corresponding to a physical state is the steady state $\hat{F}_{1}=\hat{\rho}(\infty)$. We then have
\begin{align} \label{eq:rho_long_time}
\hat{\rho}(\tau) \approx \hat{\rho}(\infty)+ \sum_{\mu=2}^{d^2}e^{\lambda_{\mu}(\tau-\tauL)}\textrm{Tr}(\hat{G}_{\mu}\hat{\mathcal{S}}[\hat{\rho}(0)])\hat{F}_{\mu},
\end{align}
where $d$ is the dimension of the system $S$ Hilbert space. The ordering of eigenvalues is given by $|\textrm{Re}(\lambda_{2})| \leq |\textrm{Re}(\lambda_{3})| \leq ... \leq |\textrm{Re}(\lambda_{d^2})|$, so that the timescale for the {\em slowest} relaxation is given by $\tau_{\textrm{re}} = 1/|\textrm{Re}(\lambda_{2})|$.
The relaxation process is then determined by the decay mode components of the state after the slippage, $\alpha_{\mu}(\rho(0)) = \textrm{Tr}(\hat{G}_{\mu}\hat{\mathcal{S}}[\hat{\rho}(0)])$. In general, $[\hat{\mathcal{S}},\hat{\mathcal{L}}_{\rm m}] \neq 0$, meaning $\hat{\mathcal{S}}$ will perturb $\hat{\rho}(\infty)$ itself, a property unique to non-Markovian systems. Since NMQMpE is defined by a steady state it applies both in and out of equilibrium. 

{\it Weak, strong and extreme NMQMpE---}
There are three possible types of NMQMpE which we call weak, strong and extreme. Let $\hat{\rho}_{\rm f}$ be some faster relaxing initial state such that $\alpha_{\nu_{0}}(\hat{\rho}_{\rm f})$ and $\alpha_{\kappa_{0}}(\hat{\rho}(\infty))$ are the components of the slowest non-zero modes excited by $\hat{\mathcal{S}}$ when starting with $\hat{\rho}_{\rm f}$ and $\hat{\rho}(\infty)$, respectively. The weak NMQMpE arises when $\nu_{0} = \kappa_{0}$ and $\alpha_{\nu_{0}}(\hat{\rho}_{\rm f})<\alpha_{\kappa_{0}}(\hat{\rho}(\infty))$. This means the slowest non-zero modes excited by $\hat{\mathcal{S}}$ are the same for both initial conditions, but $\hat{\rho}_{\rm f}$ has a smaller amplitude in that mode. Then, $\hat{\rho}_{\rm f}$ will relax faster, but not exponentially faster. If $\nu_{0}>\kappa_{0}$, there exists a strong NMQMpE where $\hat{\rho_{\rm f}}$ relaxes exponentially faster than $\hat{\rho}(\infty)$ by a rate given by the spectral gap $\textrm{Re}|\lambda_{\nu_{0}}-\lambda_{\kappa_{0}}|$. The weak and strong NMQMpE are analogs of weak and strong MpE observed in Markovian quantum dynamics \cite{PhysRevLett.127.060401,PhysRevLett.131.080402,PhysRevE.108.014130}. However, for NMQMpE, we can have an additional extreme case
if $\hat{\alpha}_{\nu}(\hat{\rho}_{\rm f}) = 0$ for all $\nu$ in which $\hat{\mathcal{S}}$ removes all decay components from $\hat{\rho}_{\rm f}$ to give the steady state $\hat{\rho}(\tau>\tauL) = \hat{\mathcal{S}}[\hat{\rho}_{\rm f}] = \hat{\rho}(\infty)$. Thus, the initial state converges to the steady state within the memory time $\tauL$, giving the fastest possible relaxation. Formally,
\begin{equation} \label{eq:rho_f}
\hat{\rho}_{\rm f} = \frac{\hat{\mathcal{S}}^{-1}[\hat{\rho}(\infty)]}{\textrm{Tr}(\hat{\mathcal{S}}^{-1}[\hat{\rho}(\infty)])}.
\end{equation}
The above state may not be physically valid in general because (i) $\mathcal{S}^{-1}$ may not exist and (ii) the resulting state may not be positive semidefinite. However, in practice these conditions don't pose an issue. First, $\hat{\Lambda}^{-1}(\tau)$ may be undefined only at some isolated time points \cite{Reimer2019Jul,nestmann2019timeconvolutionless,li2012timeconvolutionless,
chruscinski2010non}, such that we are free to shift $\tauL$ slightly to give a well defined $\mathcal{S}^{-1}$ \cite{PhysRevA.104.062403}. Second, if the system-bath coupling is not too strong, $\mathcal{S}^{-1}$ can be found perturbatively \cite{gaspard1999slippage,Purkayastha_2016} and the hermiticity of $\hat{\rho}_{\rm f}$ is guaranteed \cite{suppmat}. In the following, we show the possibility of the extreme NMQMpE in quantum dot setups and demonstrate that $\hat{\rho}_{\rm f}$ is positive semidefinite over a wide range of parameters. 

\textit{Baths for numerical examples---} We consider a system $S$ coupled to two electronic thermal reservoirs, allowing for both equilibrium and non-equilibrium dynamics, see Fig.~\ref{fig1}(a). We take the baths to be a continuum of modes, with
$
\hat{H}_{B}+\hat{H}_{SB} = \sum_{\alpha=L,R}\big(\int_{-D}^{D}\omega\hat{c}_{\alpha}^{\dag}(\omega)\hat{c}_{\alpha}(\omega)d\omega + \int_{-D}^{D}\sqrt{\mathcal{J}_{\alpha}(\omega)}(\hat{c}_{\alpha}^{\dag}(\omega)\hat{Q}_{\alpha}+\hat{Q}^{\dag}_{\alpha}\hat{c}_{\alpha}(\omega))d\omega\big),
$
where $\mathcal{J}_{\alpha}(\omega)$ is the spectral density of the bath, $\hat{Q}_{\alpha}$ is the system operator coupling to the bath and $\hat{c}_{\alpha}^{\dag}(\omega),\hat{c}_{\alpha}(\omega)$ are canonical fermionic creation and annihilation operators for the bath modes obeying $\{\hat{c}_{\alpha}^{\dag}(\omega),\hat{c}_{\alpha}(\omega')\}=\delta(\omega-\omega')$, with $\alpha=L,R$ corresponding to left and right baths. We parameterise $\mathcal{J}_{\alpha}(\omega)$ via the total coupling strength $\Gamma_{\alpha}$ defined as $\Gamma_{\alpha}/D = \frac{1}{2}\int_{-D}^{D}2\pi\mathcal{J}_{\alpha}(\omega)d\omega$, where $D$ is its bandwidth. For simplicity, we assume a semi-elliptical spectral function for both baths  $\mathcal{J}_{\alpha}(\omega) = (2\Gamma_{\alpha}/\pi^2)\sqrt{1-(\omega/D)^2}$~\footnote{Note that the NMQME is not dependent on this choice, we have confirmed it for a flat and Lorentzian band also.}, and an initial state $\hat{\rho}_{B} = \prod_{\alpha}\hat{\rho}_{\alpha}$, where $\hat{\rho}_{\alpha} = e^{-\beta_{\alpha}(\hat{H}_{\alpha}-\mu_{\alpha}\hat{N}_{\alpha})}/Z_{\alpha}$ is a thermal state and $Z_{\alpha}$ is the partition function for bath $\alpha$.

{\it Numerical methods---}
For our calculations the bath modes are discretized using a chain mapping \cite{chin2011chain,2020arXiv200403970M} combined with the thermofield transformation \cite{takahashi1996thermo,borrelli2021finite} to encode the finite temperature effects exactly over a finite time. In the absence of interactions, the dynamics can be solved exactly using unitary evolution of the single-particle correlation matrix. For interacting systems, where $\hat{H}$ is no longer quadratic, we instead use the time dependent variational principle \cite{Yang_2020,Fishman_2022,Haegeman_2016} applied to matrix product states~\cite{Fishman_2022} to simulate the real-time dynamics of the system and baths. We extract $\hat{\mathcal{S}}$ from the evolved system state via the Choi-Jamiolkowski isomorphism~\cite{suppmat}. 
 \begin{figure}[t]  
    \centering
    \includegraphics[width=0.5\textwidth]{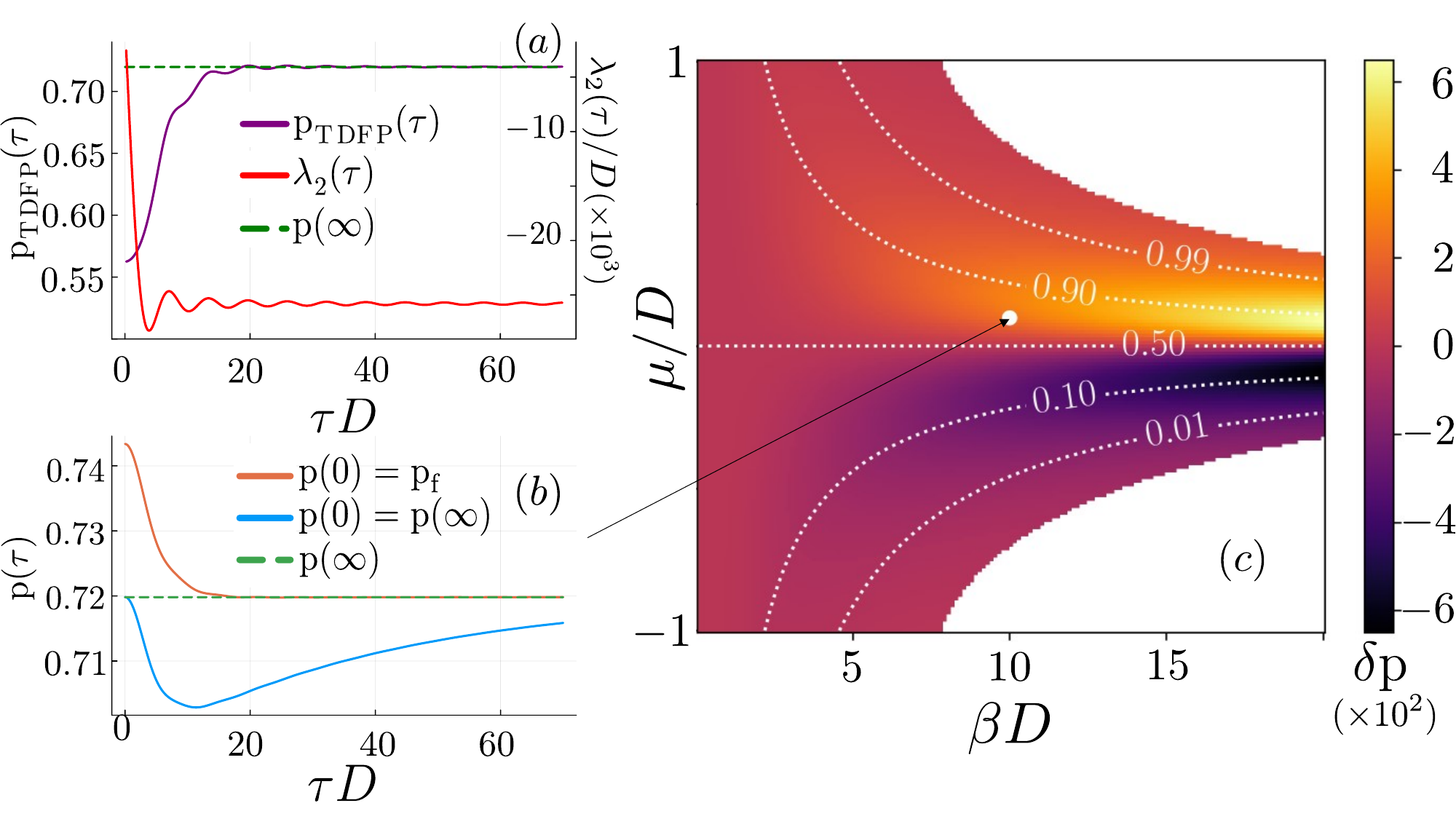}
    \caption{NMQMpE for a single quantum dot. (a) Time-dependent fixed point population ${\rm p}_{\rm TDFP}(\tau)$ (left axis) and relaxation rate $\lambda_2(\tau)$ (right axis). (b) Relaxation of population of the QD towards steady state, when starting from  $\textrm{p}(0)=\textrm{p}_{\rm f}$ and $\textrm{p}(0)=\textrm{p}(\infty)$. The dashed lines in panels (a) and (b) correspond to $\textrm{p}(\infty)$. (c) Plot of $\delta \textrm{p} \equiv \textrm{p}_{\rm f}-\textrm{p}(\infty)$ (color-coded) as function of $\beta, \mu$, with contours of constant $\textrm{p}(\infty)$  overlaid. The white represents regions with no extreme NMQMpE. Parameters: $\varepsilon=0,\beta=10/D,\mu=0.1D,\Gamma=0.01D.$}
    \label{fig:Single mode}
\end{figure}
\par
\textit{Quantum dot---} We first demonstrate our result for a quantum dot (QD) described by a single fermionic mode $\hat{s}$ with an energy $\varepsilon$, giving $\hat{H}_{S} = \varepsilon\hat{s}^{\dag}\hat{s}$, and coupled via $\hat{Q}_\alpha = \hat{s}$ to a single bath with inverse temperature $\beta$ and chemical potential $\mu$. In this case, the state space is completely parameterized by the population $\hat{\rho}(\tau)=\hat{\rho}[\textrm{p}(\tau)] = [1-\textrm{p}(\tau)]\ket{0}\bra{0} + \textrm{p}(\tau)\ket{1}\bra{1}$ with $0 \leq \textrm{p}(\tau) \leq 1$ being the population of the QD,with $\ket{1}=\hat{s}^{\dag}\ket{0}$. For the QD, $\hat{\Lambda}(\tau)$ and $\hat{\mathcal{L}}(\tau)$ can be written as $2\times 2$ matrices. One eigenvalue of $\hat{\mathcal{L}}(\tau)$ is always $0$. The corresponding eigenvector gives the time-dependent-fixed-point, $\hat{\rho}_{\rm TDFP}(\tau)$, i.e, $\hat{\mathcal{L}}(\tau)[\hat{\rho}_{\rm TDFP}(\tau)] = 0$. We denote the corresponding population ${\rm p}_{\rm TDFP}(\tau)$. The other eigenvalue of $\hat{\mathcal{L}}(\tau)$, which we denote  $\lambda_2(\tau)$, gives the instantaneous decay rate. In Fig \ref{fig:Single mode}(a), we plot ${\rm p}_{\rm TDFP}(\tau)$ and $\lambda_2(\tau)$ along with ${\rm p}(\infty)$ obtained exactly from non-equilibrium Green's functions \cite{mrsbulletin_2017,Purkayastha_2019,cornean2005rigorous}. We see that ${\rm p}_{\rm TDFP}(\tau)$ saturates to ${\rm p}(\infty)$ and $\lambda_2(\tau)$ becomes approximately constant in a time $\tau \approx 20/D$ which we take as $\tauL$. Given $\tauL$ and ${\rm p}(\infty)$, the population $\textrm{p}_{\rm f}$ of $\hat{\rho}_{\rm f}$ [Eq.\eqref{eq:rho_f}] is \cite{suppmat}
\begin{equation}
\textrm{p}_{\rm f} = \frac{\bra{0}\hat{\mathcal{S}}[\ket{0}\bra{0}]\ket{0}-[1-\textrm{p}(\infty)]}{\bra{0}\hat{\mathcal{S}}[\ket{0}\bra{0}-\ket{1}\bra{1}]\ket{0}}.
\end{equation}
In Fig.~\ref{fig:Single mode}(b), We clearly see that when starting from  ${\rm p}(0)={\rm p}(\infty)$, the population is quickly perturbed away before slowly relaxing back. In contrast, when starting from ${\rm p}(0)={\rm p}_{\rm f}$, the population relaxes to ${\rm p}(\infty)$ in time $\tauL$. Thus, we demonstrate the extreme NMQMpE in the QD. Fig.~\ref{fig:Single mode}(c) shows the deviation in density of $\hat{\rho}_{\rm f}$ from $\hat{\rho}(\infty)$, serving as a measure of the strength of the NMQMpE across the $(\mu,\beta)$-phase diagram. The effect disappears for high $|\mu|$, $\beta$ as $\textrm{p}(\infty) \approx 1$ for $\mu>0$ and $\textrm{p}(\infty) \approx 0$ for $\mu<0$ giving $\textrm{p}_{\rm f}>1$ and $\textrm{p}_{\rm f}<0$ respectively, shown by the white regions. The increase in $\delta {\rm p} = \textrm{p}_{\rm f}-\textrm{p}(\infty)$ for increasing $\beta$ reflects the expected increase in non-Markovian behaviour as temperature decreases. For a single quantum dot the effect is always an effective equilibrium one, since any $\hat{\rho}$ can be expressed as a thermal state.
\par
\textit{Double quantum dot---} To explore a non-equilibrium setup we consider a double quantum dot (DQD) setting described by modes $\hat{s}_1$ and $\hat{s}_2$, with
$
\hat{H}_{S} = g(\hat{s}^{\dag}_{2}\hat{s}_{1} + \hat{s}^{\dag}_{1}\hat{s}_{2})+U\hat{n}_{1}\hat{n}_2,
$
where $\hat{n}_{i}=\hat{s}^{\dag}_{i}\hat{s}_{i}$.  Each mode is connected to its own bath via $\hat{Q}_{\rm L}=\hat{s}_{1}$ and $\hat{Q}_{\rm R}=\hat{s}_{2}$ with the bath setup parameterised as $\delta\mu= (\mu_{\rm L}-\mu_{\rm R})/2$, $\bar{\mu}=(\mu_{L}+\mu_{R})/2$, $\beta_{L}=\beta_{R}=\beta$ and $\Gamma_{L}=\Gamma_{R}=\Gamma$. To measure the distance from the steady state, we use the trace distance defined by $T[\hat{\rho}(\tau),\hat{\rho}(\infty)]\equiv \frac{1}{2}\textrm{Tr}[\sqrt{(\hat{\rho}(\tau)-\hat{\rho}(\infty))^{\dag}(\hat{\rho}(\tau)-\hat{\rho}(\infty))}]$ and we also consider the dynamics of the particle current between the two systems modes, quantified by $\langle J(\tau) \rangle = i{\rm Tr}[\hat{\rho}(\tau)(\hat{s}^{\dag}_{1}\hat{s}_{2} - \hat{s}^{\dag}_{2}\hat{s}_{1})]$. 
For the DQD, $\hat{\Lambda}(\tau)$ and $\hat{\mathcal{L}}(\tau)$ are now $16 \times 16$ matrices. Similar to the QD case, we estimate $\tauL$ as the time beyond which $\hat{\rho}_{\rm TDFP}(\tau)$ becomes approximately constant, and take $\hat{\rho}(\infty)=\hat{\rho}_{\rm TDFP}(\tauL)$.
 \begin{figure}[t]  
    \centering
    \includegraphics[width=0.5\textwidth]{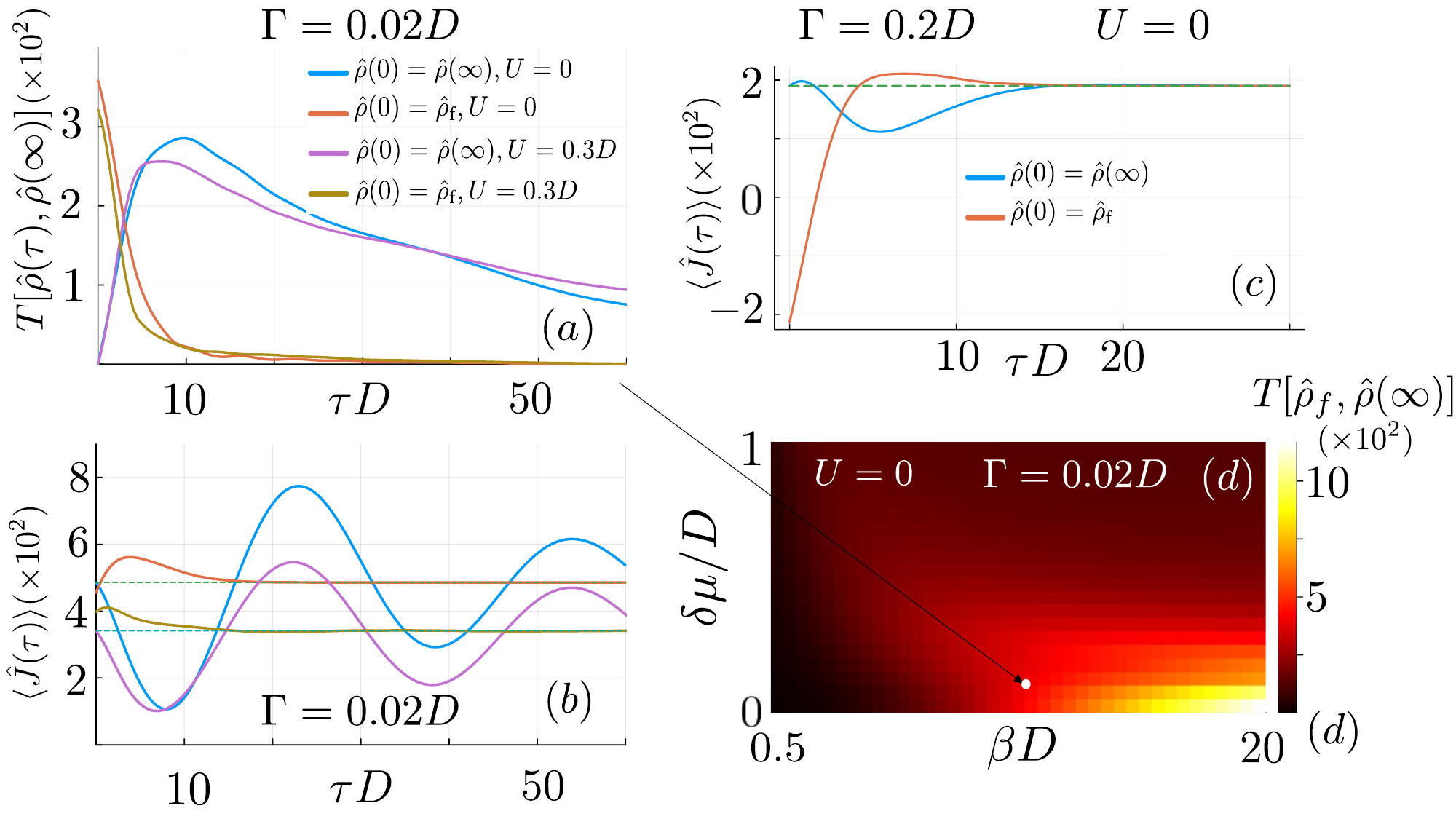}
        \caption{NMQMpE for two quantum dots in a non-equilibrium setup. Relaxation towards steady state for $\hat{\rho}(0)=\hat{\rho}_{\rm f}$ and $\hat{\rho}(0)=\hat{\rho}(\infty)$, for both non-interacting and interacting setups in terms of (a) trace distance and (b) currents with (c) showing the case of strong coupling with no interactions. The dashed lines in (b) and (c) show NESS currents. (d) Trace distance between $\hat{\rho}_{\rm f}$ and $\hat{\rho}(\infty)$ for various $\delta\mu$ and $\beta$. Parameters: $\Gamma_{\rm L}=\Gamma_{\rm R}=\Gamma,g=0.1D,\beta_{L}=\beta_{R}=10D,\bar{\mu}=0,\delta\mu=0.1D$. }
    \label{fig:Two modes}
\end{figure}
\par

Figures~\ref{fig:Two modes}(a),(b) show the NMQMpE for weak coupling with and without interactions. In both cases, there is a clear NMQMpE as can be seen from the crossing of $T[\hat{\rho}_{\textrm{f}}(\tau),\hat{\rho}(\infty)]$ with $T[\hat{\rho}_{\textrm{SS}}(\tau),\hat{\rho}(\infty)]$ in Fig.~\ref{fig:Two modes}(a), a classic characteristic of an MpE. Figure~\ref{fig:Two modes}(b) reports the associated currents, where $\hat{\rho}(\infty)$ displays a damped large amplitude oscillation with and without interactions, consistent with the slowest decaying mode with $\tau_{\rm re}=1/|\rm Re(\lambda_{2})|$, while $\hat{\rho}_{\rm f}$ shows no such oscillation.

In Fig.~\ref{fig:Two modes}(c) we consider a stronger coupling. In this case the slippage causes a large perturbation resulting in the initial current for $\hat{\rho}(0)=\hat{\rho}_{\rm f}$ being reversed from that of $\hat{\rho}(\infty)$. However, the NMQMpE is less pronounced in the subsequent time evolution because stronger coupling substantially reduces the difference in timescales $\tauL - \tau^{\Lambda}_{\rm m}$ \cite{suppmat}. 

Concentrating on the weak coupling regime, Fig.~\ref{fig:Two modes}(d) shows the trace distance $T[\hat{\rho}_{\rm f},\hat{\rho}(\infty)]$ for a range of $\delta\mu$ and non-zero $\beta$ with $\bar{\mu}=0$. In contrast to Fig.~\ref{fig:Single mode}(c), here $\hat{\rho}_{\rm f}$ has a physical solution for the entire parameter regime as $\hat{\rho}(\infty)$ lies far from the boundary of physical states \cite{suppmat}.
Figure~\ref{fig:Two modes}(d) displays increasing $T[\hat{\rho}_{\rm f},\hat{\rho}(\infty)]$ with lower temperature consistent with the dynamics being more non-Markovian. A surprising feature of Fig.~\ref{fig:Two modes}(d) is that the greatest deviation occurs at equilibrium $\delta\mu= 0$ as $\bar{\beta}\to\infty$ where both modes are always half filled with zero current. Here $\hat{\rho}(\tau)$ can only differ by $\textrm{Re}(\langle\hat{s}_1^{\dag}\hat{s}_2\rangle)$, such that the relaxation dynamics is entirely determined by the quantum coherence.

\textit{Possible experimental verification---} The QD setup lends itself to experimental verification of the NMQMpE by measuring the dynamics of the dot's occupation using a quantum point contact charge sensor. Assuming a measurement time resolution of $\approx 10~\mu {\rm s}$ \cite{Elzerman_2004} requires a memory time $\tauL\geq20~\mu {\rm s}$ for the slippage to be observable. Further, assuming an occupation resolution of $1\%$ for the measurement then requires $\delta\rm p \geq 0.01$ so that $\hat{\rho}_{\rm f}$ can be reliably distinguished from $\hat{\rho}(\infty)$. This is satisfied for the case shown in Fig.~\ref{fig:Single mode}(a) for a narrow bandwidth bath $D = 1~\mu {\rm s}^{-1}$ implying a low temperature $T \approx 5~\mu {\rm K}$. In an realistic setup the control of dot and bath couplings will not be a sudden quench. This poses no issue for NMQMpE as finite time quenches can simply be incorporated into the slippage $\hat{\mathcal{S}}$ and thus accounted for in $\hat{\rho}_{\rm f}\propto\hat{\mathcal{S}}^{-1}[\hat{\rho}(\infty)]$~\cite{suppmat}.

\textit{Conclusion---} The NMQMpE uncovered in this paper are quite generic effects in non-Markovian dynamics, with no parallel in Markovian dynamics, and are accessible to experimental verification. We anticipate that NMQMpE will find applications in control of open quantum systems, for example, in quantum state preparation and qubit reset \cite{Koch_2022, Koch_2016}. For a single bath in step 2 of the process in Fig. \ref{fig1}(a), the extreme NMQMpE provides the quickest shortcut to equilibriation \cite{Dann_2019, Boubakour_2024},  while in the presence of multiple baths, it allows the quickest steady state preparation. Such steady states can have quantum coherence \cite{Guarnieri_2018, Purkayastha_2020, Cresser_2021, Trushechkin_2022} and correlations \cite{Tavakoli_2020,Bohr_Brask_2022} which may be exploited in quantum technologies. Furthermore, NMQMpE may influence the performance of finite-time cyclic quantum thermal machines \cite{Bhattacharjee_2021,PhysRevResearch.2.043247,Liu_2021,Mukherjee_2020,Purkayastha_2022,Khait_2022}.  Finally, there may be fundamental connections between NMQMpE and quantum speed limits in dissipative systems \cite{del_Campo_2013,Liu_2016,Meng_2015,Funo_2019,Oconnor_2021,Garcia-Pintos_2022,Lan_2022}. Detailed investigations in these directions will be carried out in future works.

\begin{acknowledgments}
D.J.S and S.R.C. acknowledge insightful discussions with Francesco Turci, John Goold and Tony Short. S.R.C. also gratefully acknowledges financial support from UK's Engineering and Physical Sciences Research Council (EPSRC) under grant EP/T028424/1.
\end{acknowledgments}
\clearpage{}
\bibliography{MPE_References.bib}
\clearpage

\onecolumngrid
\begin{center}
  \textbf{\large Supplementary Material for: Non-Markovian Quantum Mpemba effect}\\[.2cm]
  David J. Strachan,$^{1}$ Archak Purkayastha,$^{2}$ and Stephen R. Clark$^1$\\[.1cm]
  {\itshape ${}^1$H. H. Wills Physics Laboratory, University of Bristol, Bristol BS8 1TL, United Kingdom\\
  ${}^2$Department of Physics, Indian Institute of Technology, Hyderabad 502284, India\\}
(Dated: \today)\\[1cm]
\end{center}

\setcounter{equation}{0}
\setcounter{figure}{0}
\setcounter{table}{0}

\renewcommand{\theequation}{S\arabic{equation}}
\renewcommand{\thefigure}{S\arabic{figure}}
\renewcommand{\bibnumfmt}[1]{[S#1]}
\renewcommand{\citenumfont}[1]{S#1}

\section{Redfield Equation}
Here we show the slippage $\hat{\mathcal{S}}$ is always invertible in the weak coupling regime. Consider a Hamiltonian $\hat{H} = \hat{H}_{S} +\hat{H}_{B}+\gamma\hat{H}_{SB}$ with $\hat{H}_{SB} = \sum_{\alpha}\hat{Q}_{\alpha}\hat{B}_{\alpha}^{\dag}+{\rm h.c.}$, $\hat{H}_{B} = \sum_{\alpha}\hat{H}_{B_{\alpha}}$ where the sum over $\alpha$ includes all baths in contact with the system and $||\hat{H}_{SB}|| = O(1)$. The Redfield equation is then the second order approximation to the Nakajima-Zwanzig master equation \cite{Purkayastha_2016,10.1093/acprof:oso/9780199213900.001.0001},
\begin{align} \label{eq:10}
\frac{d\hat{\rho}}{d\tau} &= i[\hat{\rho},\hat{H}] -\gamma^{2}\sum_{\alpha}([\hat{Q}^{\dag}_{\alpha},\hat{Q}^{1}_\alpha(\tau)\,\hat{\rho}] + [\hat{\rho}\,\hat{Q}^{2}_{\alpha}(\tau),\hat{Q}^{\dag}_\alpha]+ {\rm h.c.}), \nonumber \\
&= \hat{\mathcal{L}}_{\rm RE}(\tau)\hat{\rho}(\tau),
\end{align}
with
\begin{alignat}{2}  \label{eq:11}
\hat{Q}^{1}_{\alpha}(\tau) = \int_{0}^{\tau}d\tau'\hat{Q}^{I}_{\alpha}(-\tau')\langle\hat{B}_{\alpha}(0)\hat{B}^{I^{\dag}}_{\alpha}(-\tau')\rangle_{B},
\quad&
    \quad &
\hat{Q}^{2}_{\alpha}(\tau) = \int_{0}^{\tau}d\tau'\hat{Q}^{I}_{\alpha}(-\tau')\langle\hat{B}^{I}_{\alpha}(-\tau')\hat{B}^{\dag}_{\alpha}(0)\rangle_{B},
\end{alignat}
where $\hat{Q}^{I}_\alpha(\tau) = e^{i\hat{H}_{S}\tau}\hat{Q}_\alpha e^{-i\hat{H}_{S}\tau}$, $\hat{B}^{I}_\alpha(\tau) = e^{i\hat{H}_{B_{\alpha}}\tau}\hat{B}_\alpha e^{-i\hat{H}_{B_{\alpha}}\tau}$ and the average is taken over bath $\alpha$. This description works well when $\gamma$ is small compared to the other energy scales present. The memory timescale is then given by the time such that 
\begin{align}  \label{eq:12}
|\langle\hat{B}_{\alpha}(0)\hat{B}^{I^{\dag}}_{\alpha}(-\tau')\rangle_{B}| < O(\epsilon) \quad \forall \tau>\tau^{\mathcal{L}}_{\rm m}, \nonumber \\
|\langle\hat{B}^{I}_{\alpha}(-\tau')\hat{B}^{\dag}_{\alpha}(0)\rangle_{B}| < O(\epsilon) \quad \forall \tau>\tau^{\mathcal{L}}_{\rm m},
\end{align}
where $\epsilon$ is an arbitrarily small error. This then gives a time independent generator for $\tau>\tau_{\rm m}^{\mathcal{L}}$
\begin{align} \label{eq:13}
\frac{d\hat{\rho}}{d\tau} &= i[\hat{\rho},\hat{H}] -\gamma^{2}\sum_{\alpha}([\hat{Q}^{\dag}_{\alpha},\hat{Q}^{1}_\alpha\,\hat{\rho}] + [\hat{\rho}\,\hat{Q}^{2}_{\alpha},\hat{Q}^{\dag}_\alpha]+{\rm h.c.}),
\end{align}
with
\begin{alignat}{2}  \label{eq:14}
\hat{Q}^{1}_{\alpha} = \int_{0}^{\infty}d\tau'\hat{Q}^{I}_{\alpha}(-\tau')\langle\hat{B}_{\alpha}(0)\hat{B}^{I^{\dag}}_{\alpha}(-\tau')\rangle_{B},
\quad&
    \quad &
\hat{Q}^{2}_{\alpha} = \int_{0}^{\infty}d\tau'\hat{Q}^{I}_{\alpha}(-\tau')\langle\hat{B}^{I}_{\alpha}(-\tau')\hat{B}^{\dag}_{\alpha}(0)\rangle_{B}.
\end{alignat}
We now calculate the fast state $\hat{\rho}_{\rm f}=\hat{\mathcal{S}}^{-1}[\hat{\rho}_{\rm SS}$]. Solving Eq.(\ref{eq:13}) up to $O(\gamma^{2})$ gives 
\begin{equation}  \label{eq:15}
\hat{\rho}(\tau) = e^{-i\hat{H}_{S}(\tau)}\hat{\rho}(0)e^{i\hat{H}_{S}(\tau)} - \gamma^{2}\sum_{\alpha}\int_{0}^{\tau}d\tau'e^{-i\hat{H}_{S}(\tau-\tau')}\bigg([\hat{Q}^{\dag}_{\alpha},\hat{Q}^{1}_\alpha(\tau')\,\hat{\rho}(\tau')] + [\hat{\rho}(\tau')\,\hat{Q}^{2}_{\alpha}(\tau'),\hat{Q}^{\dag}_\alpha]+{\rm h.c.}\bigg)e^{i\hat{H}_{S}(\tau-\tau')},
\end{equation}
and hence
\begin{equation}  \label{eq:16}
\hat{\rho}(0) = e^{i\hat{H}_{S}\tau}\hat{\rho}(\tau)e^{-i\hat{H}_{S}\tau} - \gamma^{2}\sum_{\alpha}\int_{0}^{\tau}d\tau'e^{i\hat{H}_{S}\tau'}\bigg([\hat{Q}^{\dag}_{\alpha},\hat{Q}^{1}_\alpha(\tau')\,\hat{\rho}] + [\hat{\rho}\,\hat{Q}^{2}_{\alpha}(\tau'),\hat{Q}^{\dag}_\alpha]+{\rm h.c.}\bigg)e^{-i\hat{H}_{S}\tau'},
\end{equation}
with $\hat{\rho}(\tau') = e^{-i\hat{H}_{S}(\tau'-\tau)}\hat{\rho}(\tau)e^{i\hat{H}_{S}(\tau'-\tau)} +O(\gamma^{2})$. Setting $\tau=\tau^{\mathcal L}_{\rm m}$ and $\hat{\rho}(\tau^{\mathcal L}_{\rm m})=\hat{\rho}_{\rm SS}$, we find the $\hat{\rho}_{\rm f}$ up to $O(\gamma^{2})$ 
\begin{align}  \label{eq:17}
\hat{\rho}_{\rm f} &=\hat{\rho}(0) = e^{i\hat{H}_{S}\tau}\hat{\rho}(\tau)e^{-i\hat{H}_{S}\tau} \nonumber \\
&- \gamma^{2}\sum_{\alpha}\int_{0}^{\tau}d\tau'e^{i\hat{H}_{S}\tau'}\bigg([\hat{Q}^{\dag}_{\alpha},\hat{Q}^{1}_\alpha(\tau')e^{-i\hat{H}_{S}(\tau'-\tau)}\hat{\rho}(\tau)e^{i\hat{H}_{S}(\tau'-\tau)}] + [e^{-i\hat{H}_{S}(\tau'-\tau)}\hat{\rho}(\tau)e^{i\hat{H}_{S}(\tau'-\tau)}\hat{Q}^{2}_{\alpha}(\tau'),\hat{Q}^{\dag}_\alpha]+{\rm h.c.}\bigg)e^{-i\hat{H}_{S}\tau'}.
\end{align}
This formulation shows $\hat{\rho}_{\rm f}$ is always defined in the weak coupling regime. This state may not be physical however, as the Redfield description preserves hermicity and trace but not positivity. If this solution gives non-negative eigenvalues for $\hat{\rho}_{\rm f}$, there exists a NMQMpE.

\section{Calculation of $\hat{\Lambda}(\tau)$} \label{appendix: phase gates}
To extract $\hat{\Lambda}(\tau)$, we use the Choi-Jamiolkowski isomorphism \cite{CHOI1975285,JAMIOLKOWSKI1972275} which describes the correspondence between quantum maps and quantum states. Consider maximally entangling the system with an auxillary system $A$ with the same Hilbert space $\mathcal{H}_{S}$ as $S$,  $\ket{\Phi^{+}}=\frac{1}{\sqrt{d}}\sum_{i=0}^{d-1}\ket{i}_{S}\otimes\ket{i}_{A}$. Since $\hat{\Lambda}(\tau)$ is a completely positive trace preserving map, $\hat{\rho}^{\Lambda}(\tau) = (\hat{\Lambda}(\tau)\otimes\mathds{1})\{\ket{\Psi^{+}}\bra{\Psi^{+}}\}$ is a nonnegative operator. Conversely, for any nonnegative operator on $L(\mathcal{H}_{S})\otimes L(\mathcal{H}_{S})$, we can associate a quantum map from operators on $L(\mathcal{H}_{S})$ to $L(\mathcal{H}_{S})$. This isomorphism assumes a tensor product Hilbert space structure and can be applied to 1D fermionic systems once they have been Jordan Wigner transformed into effective spins. The fermionic map can then be extracted once the anti-commuting behaviour of fermions is accounted for \cite{Amosov_2016}.

\section{Time evolution and Bath description} \label{appendix:time evolution}
For numerical calculations the continuum of modes for each bath $\alpha$ is approximated using a finite number of modes $N_{\alpha}$. We use the same number of modes for each bath $N_{\alpha} = N_{b}$. To do this we employ a finite chain mapping using orthogonal polynomials and a thermofield purification scheme to describe the finite temperature bath initial states. This results in the following Hamiltonian,
\begin{align} 
\hat{H} &= \hat{H}_{S} +\sum_{\tilde{\alpha}}\bigg[\kappa_{\tilde{\alpha},0}^{-1}(\hat{s}^{\dag}\hat{b}_{\tilde{\alpha},0}+\hat{b}^{\dag}_{\tilde{\alpha},0}\hat{s}) +\sum_{n=0}^{N_{b}}\bigg( \gamma_{\tilde{\alpha},n}\hat{b}^{\dag}_{\tilde{\alpha},n}\hat{b}_{\tilde{\alpha},n} 
+\sqrt{\beta_{\tilde{\alpha},n+1}}\hat{b}^{\dag}_{\tilde{\alpha},n}\hat{b}_{\tilde{\alpha},n+1} +\sqrt{\beta_{\tilde{\alpha},n+1}}\hat{b}^{\dag}_{\tilde{\alpha},n+1}\hat{b}_{\tilde{\alpha},n}\bigg)\bigg].  \label{eq:full_ham}
\end{align}
Here $\tilde{\alpha} = L_{f},L_{e},R_{f},R_{e}$ now denotes the four baths, left and right correspondingly filled and empty, with all their parameters fully defined by the bath spectral functions, temperatures and chemical potentials. The 1D mode ordering used is given by 
\begin{align}  \label{eq:19}
\{\hat{o}_{i}\} = \{\hat{b}_{L_{f}1},\hat{b}_{L_{e}1},...,\hat{b}_{L_{f}N_{b}},\hat{b}_{L_{e}N_{b}},
\hat{s}_{1},...\hat{s}_{N_{s}},\hat{a}_{1},...,\hat{a}_{N_{s}},\hat{b}_{R_{f}1},\hat{b}_{R_{e}1},...,\hat{b}_{R_{f}N_{b}},\hat{b}_{R_{e}N_{b}}\},
\end{align}
where $\hat{a}_{i}$ are the system ancilla modes. The details of this scheme is outlined in Sec.~\ref{appendix: Thermofield and chain mapping}.

If $\hat{H}_{S}$ is quadratic in mode operators, the exact dynamics can then be obtained via unitary evolution of the single-particle correlation matrix $\boldsymbol{C}_{ij}(t) = \textrm{Tr}(\hat{\rho}(t)\hat{o}^{\dag}_{j}\hat{o}_{i})$ using the quadratic Hamiltonian defined via $\hat{H} = \sum_{ij}\boldsymbol{h}_{ij}\hat{o}^{\dag}_{i}\hat{o}_{j}$ and
\begin{equation} \label{eq:20}
\boldsymbol{C}(t) = e^{i\boldsymbol{h}t}\boldsymbol{C}(t_{0})e^{-i\boldsymbol{h}t},
\end{equation}
as shown in detail in Sec.~\ref{appendix: corr propagation}. The reduced density matrix of the combined system + ancilla setup is related to $C(t)$ via \cite{cheong2004many}
\begin{equation} \label{eq:21}
\hat{\rho}^{\Lambda}(t) = \mathrm{det}(\mathbb{1}-\boldsymbol{C}^{\rm T}(t))\mathrm{exp}\bigg\{\sum_{ij \in SA}[\mathrm{log}(\boldsymbol{C}^{\rm T}(t))(\mathbb{1}-\boldsymbol{C}^{\rm T}(t))^{-1}]_{ij}\hat{o}^{\dag}_{i}\hat{o}_{j}\bigg\},
\end{equation}
where $SA$ denotes the modes spanning the system + ancilla. This equation holds provided $\hat{\rho}^{\Lambda}(t)$ is block diagonal in the number basis. For this reason, we perform a particle-hole transformation on the entangled states used in the Choi-Jamiolkowski isomorphism. If $\hat{H}_{S}$ is not quadratic, we use two site time dependent variational principle \cite{Fishman_2022} to directly obtain $\hat{\rho}^{\Lambda}(t)$.

\section{NMQMpE across coupling strength}
\begin{figure}[h!]  
    \centering
    \includegraphics[width=0.8\textwidth]{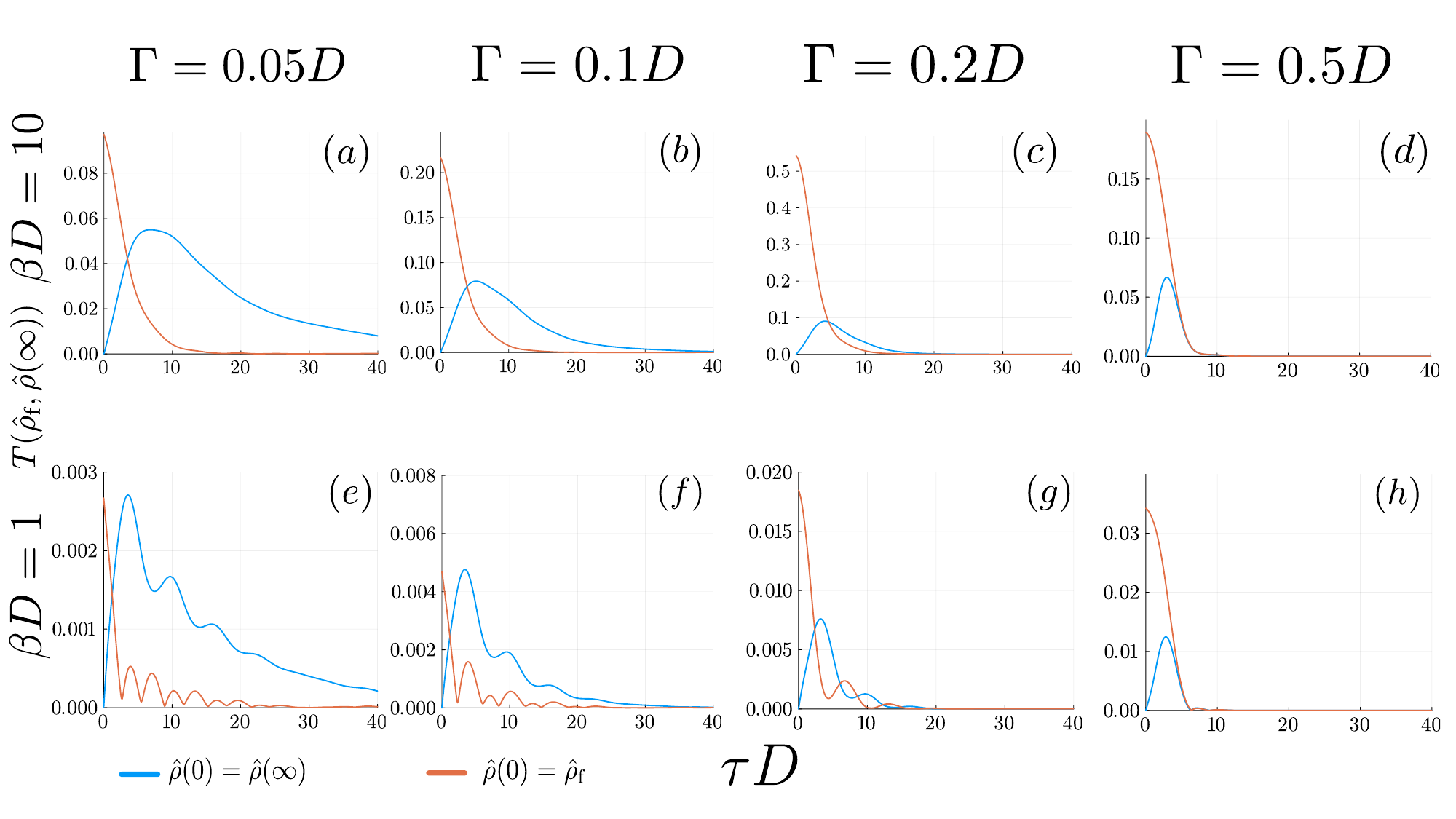}
    \caption{NMQMpE for weak, intermediate and strong couplings for two non-interacting modes in a non-equilibrium setup, $\delta\mu=0.1D,\bar{\mu}=0,\beta_{L}=\beta_{R},\Gamma_L=\Gamma_R = \Gamma, U=0,g = 0.1D$. }
    \label{fig:Various Couplings}
\end{figure}
Here in Fig.\ref{fig:Various Couplings}, we show the NMQMpE for a number of parameter regimes, displaying the effect of coupling and temperature for a non-interacting two mode setup. For each coupling strength, the colder setup $\beta D=10$ always gives a larger difference $T(\hat{\rho}_{\rm f},\hat{\rho}(\infty))$ than the hotter case $\beta D=1$, reflecting the general intuition that colder systems display more non-Markovian behaviour. For $\beta D=1$, as $\Gamma$ increases the difference in convergence $\tau^{\Lambda}_{\rm m}-\tau^{\mathcal{L}}_{\rm m}$ decreases reflecting the increased decay rates of the converged generator $\hat{\mathcal{L}}(\tau)$. In the case of strong coupling, $\tau^{\Lambda}_{\rm m}-\tau^{\mathcal{L}}_{\rm m}\to 0$ as can be seen in Figs.~\ref{fig:Various Couplings}(d) and \ref{fig:Various Couplings}(h). For $\beta D=1$, the deviation $T(\hat{\rho}_{\rm f},\hat{\rho}(\infty))$ increases for increasing $\Gamma$. This is naively expected as $\Gamma$ controls the extent to which $\hat{\mathcal{S}}$ can deviate from the identity. However, this intuition breaks down at low temperatures and strong couplings as seen in Fig.~\ref{fig:Various Couplings}(d), which shows a smaller deviation than Fig.~\ref{fig:Various Couplings}(a)-(c).

\section{Existence of NMQMpE}
\begin{figure}[h!]  
    \centering
    \includegraphics[width=0.8\textwidth]{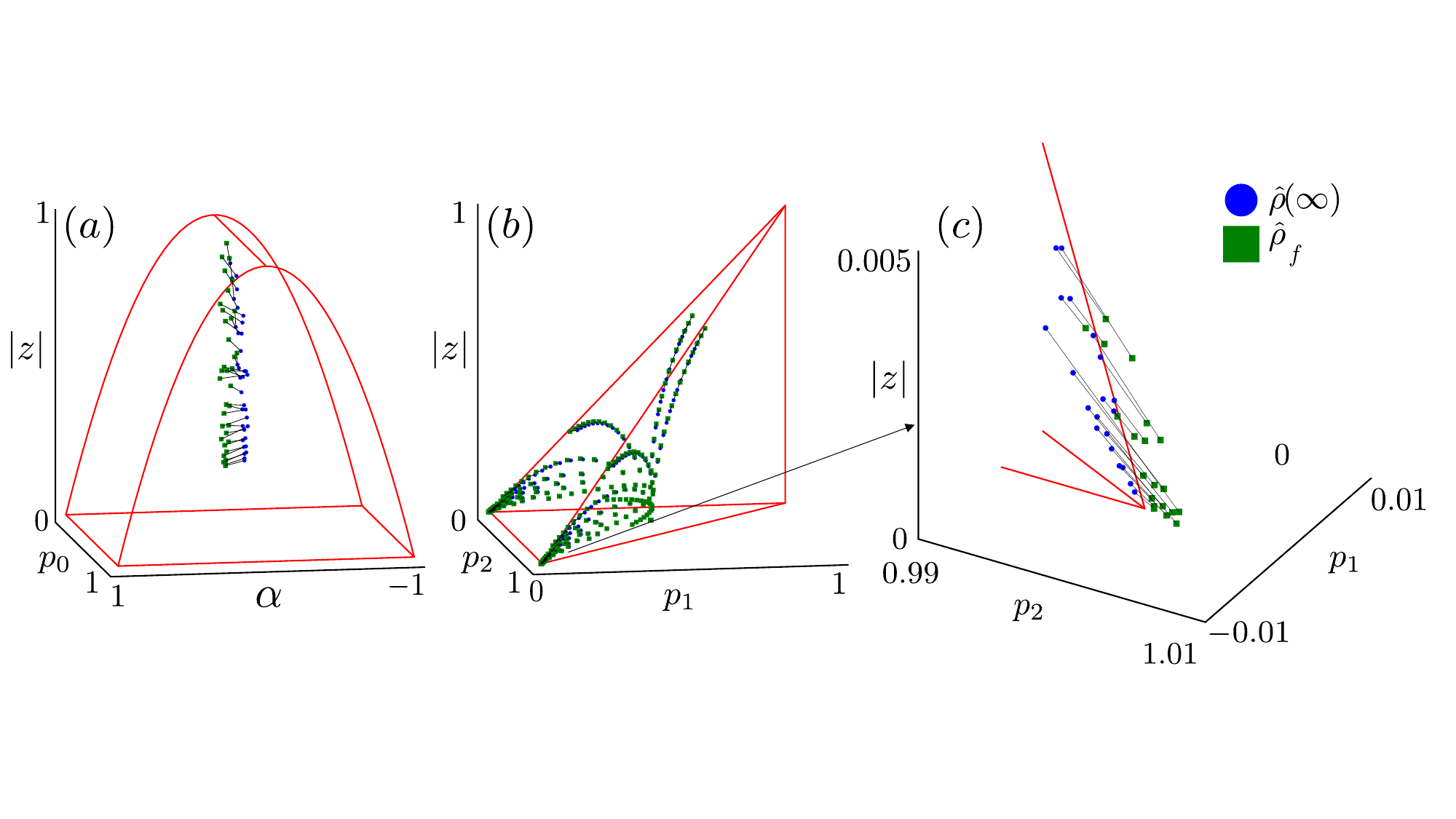}
    \caption{Parameterisation of $\hat{\rho}(\infty)$ and $\hat{\rho}_{\rm f}$ for a non-equilibrium setup (a) where $\bar{\mu}=0,0<\delta\mu<1$ and an equilibrium setup (b) $\delta\mu=0,-1<\bar{\mu}<1$. Both have equal temperature baths $\beta_{L}=\beta_R=\beta$ for $0<\beta<20$. Other parameters used: $g=0.1D,\Gamma_{L}=\Gamma_{R}=\Gamma,U=0.$}
    \label{fig:Simplex}
\end{figure}
Here we show the steady states and associated fast states in their physical spaces for the non-equilibrium setup in the main letter and an associated equilibrium setup. For the non-equilibrium setup $\bar{\mu} = 0$, $\beta_{L}=\beta_{R}$ and $\epsilon=0$, while for the equilibrium setup we have  $\delta\mu = 0$, $\beta_{L}=\beta_{R}$. The two-mode density matrices take the following general form for each of these setups 
\begin{alignat*}{2}  
\hat{\rho}_{\rm NEQ} = 
\begin{pmatrix}
p_{0} & 0 & 0 & 0 \\
0 &  p_{1}(1+\alpha)/2 & p_{1}z/2 &  0 \\
0 &  p_{1}z^{*}/2 & p_{1}(1-\alpha)/2  &  0 \\
0 & 0 & 0 & p_{0}
\end{pmatrix}, \quad&
\quad & \hat{\rho}_{\rm EQ} = 
\begin{pmatrix}
p_{0} & 0 & 0 & 0 \\
0 &  p_{1}/2 & p_{1}z/2  &  0 \\
0 &  p_{1}z^{*}/2 & p_{1}/2  &  0 \\
0 & 0 & 0 & p_{2}
\end{pmatrix},
\end{alignat*}
where $0\leq p_{i}\leq1$, $p_{1} = 1-2p_{0}$ for $\hat{\rho}_{\rm NEQ}$, while $p_{1} = 1-p_{0}-p_{2}$ for $\hat{\rho}_{\rm EQ}$ due to the trace condition and $\sqrt{\alpha^{2}+|z|^2} \leq 1$ due to positivity. In Fig.~\ref{fig:Simplex} we show this parameterisation of $\hat{\rho}_{\rm f}$ and $\hat{\rho}(\infty)$ across a range of bath temperatures and chemical potentials. Intuitively, given $\hat{\rho}_{\rm f}\propto\hat{\mathcal{S}}^{-1}[\hat{\rho}(\infty)]$ is an $O(\Gamma\tauL)$ perturbation of $\hat{\rho}_(\infty)$ then $\hat{\rho}_{\rm f}$ will be physical if it is more than $O(\Gamma\tauL)$ away from a simplex boundary. The NEQ setup gives a valid solution for $\hat{\rho}_{\rm f}$ for all parameters considered as $\hat{\rho}(\infty)$ never sits on the boundary of physically allowed states. In contrast, the EQ setup has parameter regimes with no valid $\hat{\rho}_{\rm f}$ as $\hat{\rho}(\infty)$ approaches the boundaries, as can be seen in Figs. \ref{fig:Simplex}(a) and (b). This occurs for large $|\mu|$ differences at low temperatures where the two modes are either fully empty ($p_{0}=1$), or fully occupied ($p_{2}=1)$. This issue is related to the importance of considering the slippage for the Redfield equation when starting from a state close to the boundary of physically allowed space \cite{PhysRevE.68.066112}.
\par 

\begin{figure}[h!]  
    \centering
    \includegraphics[width=0.8\textwidth]{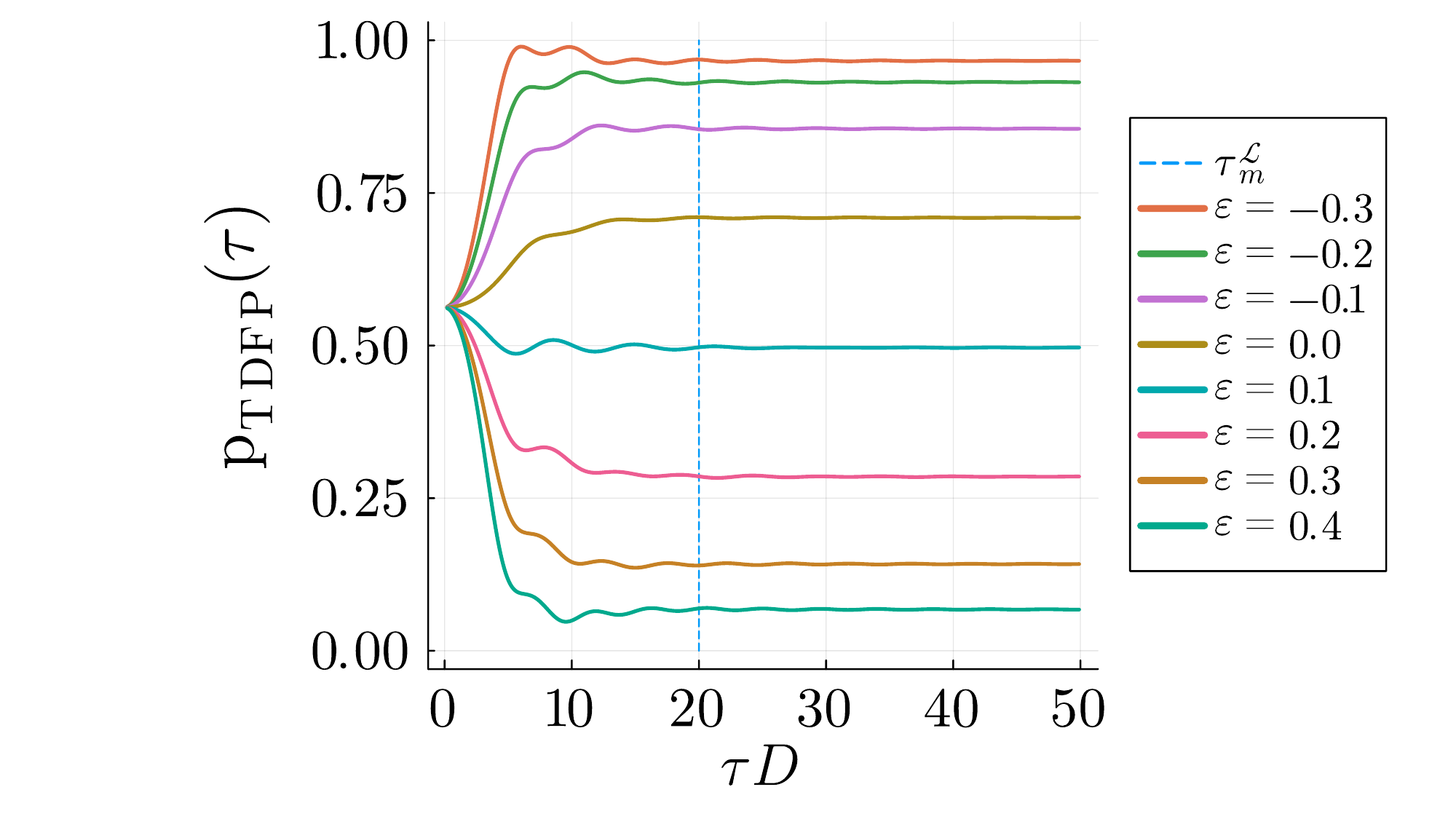}
    \caption{Convergence of $\textrm{p}_{\textrm{TDFP}}(\tau)$ for various $\varepsilon$, $\hat{H}_{S}=\varepsilon\hat{s}^{\dag}\hat{s}$. Parameters: $\beta=10/D,\mu=0.1D,\Gamma=0.01D$. }
    \label{fig:Multiple energies}
\end{figure}
Here, we demonstrate that $\tauL$ is a property of the bath only and is independent of the system parameters. To do this, we consider the convergence of $\textrm{p}_{\textrm{TDFP}}(\tau)$ for the quantum dot $\hat{H}_{S} = \varepsilon\hat{s}^{\dag}\hat{s}$, for various $\varepsilon$ as shown in Fig.~\ref{fig:Multiple energies}. From the plot, it's clear that $\varepsilon$ has no effect on the convergence of $\textrm{p}_{\textrm{TDFP}}(\tau)$ and for all system energies, convergence is approximately reached by $\tau=20/D$.
\section{Finite Quench}
\begin{figure}[h!]  
    \centering
    \includegraphics[width=0.8\textwidth]{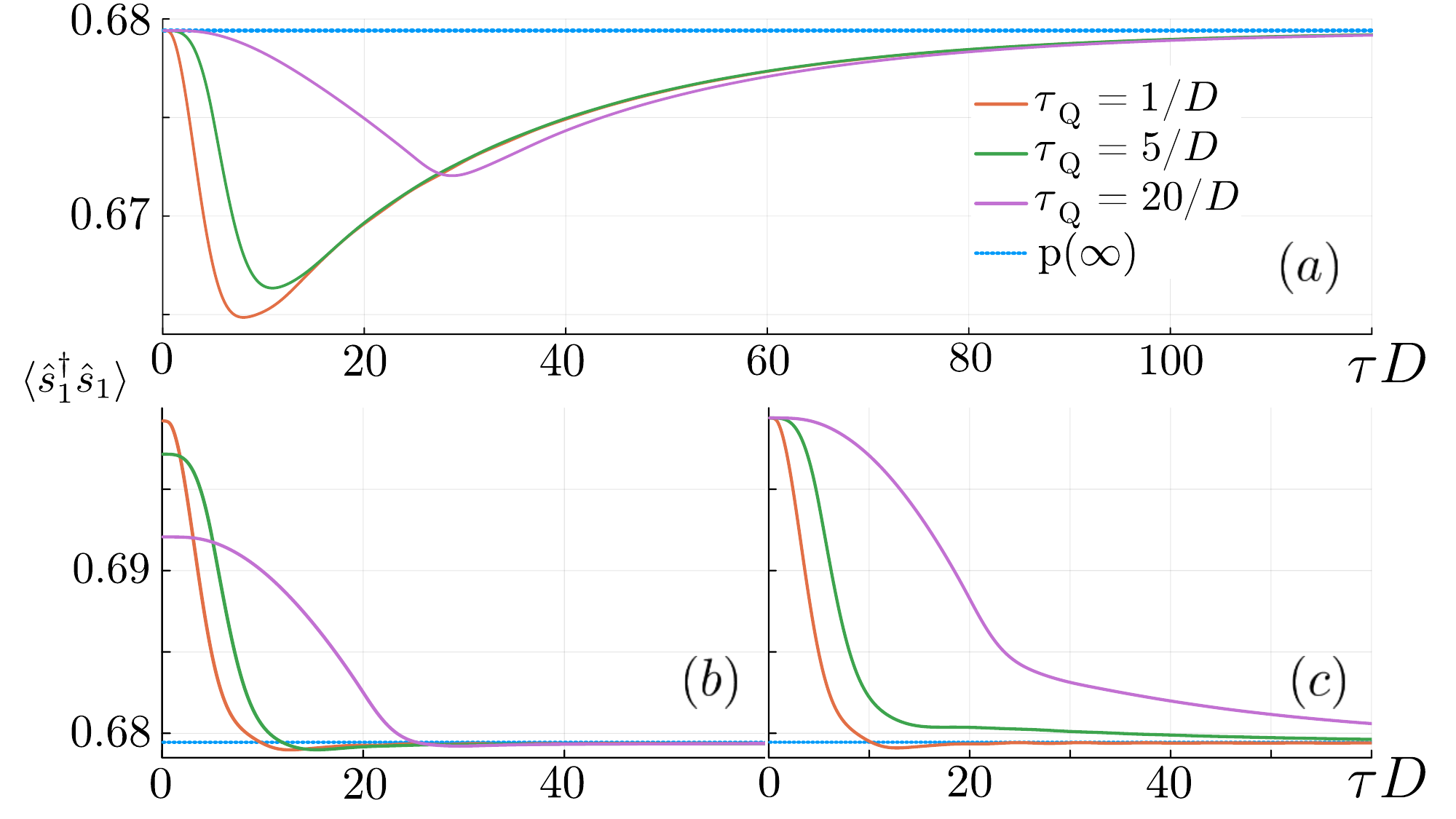}
    \caption{Simulating a finite quench for two modes using $\tau^{\mathcal L}_{\rm m} = 20D+\tau_{Q}$. (a) Convergence of $\hat{\rho}(\infty)$. (b) Convergence of $\hat{\rho}_{\rm f}(\tau_{Q})$. (c) Convergence of $\hat{\rho}_{\rm f}(\tau_{Q}=0)$. Parameters: $\epsilon=0,g=0.1D,\beta_{L}=\beta_{R}=10/D,\mu_{L}=\mu_{R}=0.1D,\Gamma_{L}=\Gamma_{R}=0.03D,U=0$.}
    \label{fig:Quench}
\end{figure}

Throughout this paper, we make the conventional assumption that the baths in step 2 and the system are disconnected until a sudden quench at $\tau=0$, i.e. $\Gamma_{\alpha}(\tau) = \Gamma_{\alpha}\Theta(\tau)$ where $\Theta(\tau)$ is the Heaviside function and $\alpha=L,R$. Realistically this will happen over a finite timescale $\tau_{Q}$ such that $\Gamma_{\alpha}(\tau) = \Gamma_{\alpha}f(\tau)$ for $\tau<\tau_{Q}$, where $f(\tau)$ continuously ramps up from $0$ at $\tau=0$ to $\Gamma_{\alpha}$ at $\tau=\tau_{Q}$. This poses no fundamental issues for NMQMpE as the existence of a state $\hat{\rho}_{\rm f}\propto\hat{\mathcal{S}}^{-1}[\hat{\rho}_{\rm SS}]$ doesn't rely on either a sudden quench or time independent bath couplings $\Gamma_{\alpha}$, and the finite quench can simply be incorporated into the slippage $\hat{\mathcal{S}}$. The robustness of NMQME to a finite $\tau_{Q}$ is shown in Fig.~\ref{fig:Quench}(b) with $f(\tau) = \Theta(\tau)\tau/\tau_{Q}$, where its clear the effect is independent of $\tau_{Q}$, if taken into account through $\hat{\Lambda}(\tau^{\mathcal L}_{\rm m})$. The existence of $\hat{\rho}_{\rm f}$ may be invariant to $\tau_{Q}$ but its form will not, i.e. $\hat{\rho}_{\rm f}$ = $\hat{\rho}_{\rm f}(\tau_{Q})$. To show this, we look at the convergence of $\hat{\rho}_{f}(\tau_{Q}=0)$ for systems with various ramp up times $\tau_{Q}$, as reported in Fig.~\ref{fig:Quench}(c). For small $\tau_{Q}$, $\hat{\rho}_{\rm f}(\tau_{Q}=0)$ remains a fast converging state but as $\tau_{Q}$ increases it deviates from the true $\hat{\rho}_{\rm f}$ and thus doesn't converge in the memory time $\tau^{\mathcal L}_{\rm m}$. Figure~\ref{fig:Quench}(a) shows the convergence of the corresponding steady state $\hat{\rho}_{\rm SS}$.
To do this, we only rescale the first bath modes of each of the four chains $L_{f},L_{e},R_{f},R_{e}$ (see Sec.~\ref{appendix: Thermofield and chain mapping}) rather than re-calculating the chain mapping at each time step in $0\leq \tau \leq \tau_{Q}$, as the rest of modes are left invariant by a rescaling of $\Gamma$. The proof is as follows. We first note that the chain mapping using orthogonal polynomials is equivalent to the reaction coordinate (RC) chain mapping. Using the reaction coordinate description for the bath chains in Eq.(\ref{eq:full_ham}) gives \cite{nano11082104,Anto_Sztrikacs_2022}
\begin{alignat}{2}
\beta_{\tilde{\alpha},n} = \int_{-D}^{D}J_{\tilde{\alpha},n}(\omega)d\omega,
\quad&
    \quad &
\gamma_{\tilde{\alpha},n} = \frac{1}{\beta_{\tilde{\alpha},n-1}}\int_{-D}^{D}J_{\tilde{\alpha},n-1}(\omega)d\omega,
\end{alignat}
where $\beta_{\tilde{\alpha},n} = \kappa_{\tilde{\alpha},0}^{-1}$ and $J_{\tilde{\alpha}}(\omega)$ is the spectral function for $\tilde{\alpha}$. The recursive spectral functions are given by 
\begin{equation} \label{eq:22}
J_{\tilde{\alpha},n+1}(\omega) = \frac{\beta_{\tilde{\alpha},n}J_{\tilde{\alpha},n}(\omega)}{\left|\pi J_{\tilde{\alpha},n}(\omega)+\mathcal{P}\int_{-D}^{D}\frac{J_{\tilde{\alpha},n}(\omega')d\omega'}{\omega'-\omega}\right|^2},
\end{equation}
where $\mathcal{P}$ denotes Cauchy's principle value. Changing the coupling strength $\Gamma\to\Gamma'$ is thus equivalent to $J_{\tilde{\alpha},0}(\omega)\to (\Gamma'/\Gamma)J_{\tilde{\alpha},0}(\omega)$. This leaves $J_{\tilde{\alpha},n}(\omega)$ invariant for $n\geq 1$ as can be seen in Eq.~\eqref{eq:22} such that only the first site in the the chain is modified.

\section{Landauer B\"{u}ttiker Theory} \label{appendix: Landauer buttiker}

Here we briefly outline how to compute the currents in Landauer-B\"{u}ttiker theory which act as our point of comparison for non-interacting system steady states. In the continuum limit for macroscopic baths, the particle and energy currents in the absence of system interactions are given by \cite{mrsbulletin_2017,Purkayastha_2019}
\begin{equation}  \label{eq:23}
J_{LB}^{P} = \frac{1}{2\pi}\int_{-D}^{D}d\omega\tau(\omega)[f_{L}(\omega)-f_{R}(\omega)],
\end{equation}
and
\begin{equation}  \label{eq:24}
J_{LB}^{E} = \frac{1}{2\pi}\int_{-D}^{D}d\omega\omega\tau(\omega)[f_{L}(\omega)-f_{R}(\omega)],
\end{equation}
where $f_{\alpha}(\omega)$ denotes the Fermi-Dirac distribution for bath $\alpha$ and $\tau(\omega)$ is the transmission function of the system. This can be calculated in terms of the non-equilibrium Green's function \cite{PhysRevB.89.165105,Purkayastha_2019}. In our case this is given by 
\begin{equation}  \label{eq:25}
G(\omega) = \boldsymbol{M}^{-1}(\omega),
\end{equation}
with
\begin{equation}  \label{eq:26}
 \boldsymbol{M}(\omega) = \omega\mathds{1}-\boldsymbol{h}_{S}-\boldsymbol{\Sigma}^{(1)}(\omega) - \boldsymbol{\Sigma}^{(N_{S})}(\omega),
\end{equation}
where the only nonzero elements of the self-energy matrices of the leads $\boldsymbol{\Sigma}^{(j)}(\omega)$ are 
\begin{equation}  \label{eq:27}
[\boldsymbol{\Sigma}^{(j)}]_{jj}(\omega) = \mathcal{P}\int_{-D}^{D}d\omega'\frac{\mathcal{J}_{j}(\omega')}{\omega'-\omega} - i\pi\mathcal{J}_{j}(\omega),
\end{equation}
using $\mathcal{J}_{1}(\omega) = \mathcal{J}_{L}(\omega)$, $\mathcal{J}_{N_{S}}(\omega) = \mathcal{J}_{R}(\omega)$ and $\boldsymbol{h}_{S}$ is defined via $\hat{H}_{S} = \sum_{ij}(\boldsymbol{h}_{S})_{ij}\hat{o}^{\dag}_{i}\hat{o}_{j}$.
If the system Hamiltonian is of the form 
\begin{equation}  \label{eq:28}
\boldsymbol{h}_{S} = \sum_{j=1}^{N_{S}}\epsilon_{j}\hat{s}^{\dag}_{j}\hat{s}_{j} +\sum_{j=1}^{N_{S}-1}t_{i}(\hat{s}^{\dag}_{j+1}\hat{s}_{j} +\text{h.c.}),
\end{equation}
the transmission function is given by

\begin{equation}  \label{eq:29}
\tau(\omega) = 4\pi^{2}\mathcal{J}_{L}(\omega)\mathcal{J}_{R}(\omega)|[G(\omega)]_{1D}|^{2} 
= \frac{\mathcal{J}_{L}(\omega)\mathcal{J}_{R}(\omega)}{|\text{det}[\boldsymbol{M}]|^2}\prod_{i=1}^{
N_{S}-1}|t_{i}|^2.
\end{equation}

\section{Single quantum dot case} \label{appendix: Single site case}
A single quantum dot is described by a spinless fermion mode as 
\begin{equation}  \label{eq:30}
\hat{H}_{S} = \varepsilon\hat{s}^{\dag}\hat{s},
\end{equation}
such that its density operator has to be block diagonal in the Fock basis $\ket{0},\ket{1}$ as
\begin{equation}  \label{eq:31}
\hat{\rho}(\textrm{p}) = (1-\textrm{p})\ket{0}\bra{0} +\textrm{p}\ket{1}\bra{1},
\end{equation}
with an occupation $0 \leq {\rm p} \leq 1$. Using $\textrm{Tr}(\hat{F}_{2}) = 0$ and $\textrm{Tr}(\hat{G}_{\mu}\hat{F}_{\nu}) = \delta_{\mu\nu}$ for a single mode we have 
\begin{alignat*}{2}
\hat{F}_{2} = \ket{0}\bra{0} -\ket{1}\bra{1},
\quad&
    \quad &
\hat{F}_{1} =  (1-\textrm{p}(\infty))\ket{0}\bra{0} +\textrm{p}(\infty)\ket{1}\bra{1},
\end{alignat*}
\begin{alignat*}{2}  \label{eq:33}
\hat{G}_{1} =  \ket{0}\bra{0} + \ket{1}\bra{1},
\quad&
    \quad &
\hat{G}_{2} = \textrm{p}(\infty)\ket{0}\bra{0} +(\textrm{p}(\infty)-1)\ket{1}\bra{1},
\end{alignat*}
where $\hat{\rho}(\infty) = \hat{\rho}(\textrm{p}(\infty))$.
Now consider the slippage $\hat{\mathcal{S}} = \hat{\Lambda}(\tauL)$. After vectorizing $\hat{\rho}$ we can express $\hat{\mathcal{S}}$ as a $ 2 \times 2$ matrix acting in the physical subspace $\{\ket{0}\bra{0},\ket{1}\bra{1}\}$. For this map to be CPTP, it must be of the form
\begin{equation}  \label{eq:34}
\hat{\mathcal{S}} = 
\begin{pmatrix}
\nu & \sigma \\
1-\nu & 1-\sigma 
\end{pmatrix}
\end{equation}
with $\nu = \bra{0}\hat{\mathcal{S}}\{\ket{0}\bra{0}\}\ket{0}$, $\sigma = \bra{0}\hat{\mathcal{S}}\{\ket{1}\bra{1}\}\ket{0}$, with $0\leq \nu, \sigma \leq 1$. Now $\hat{\mathcal{S}}^{-1}$ is manifestly trace preserving for the single mode, $\textrm{Tr}(\hat{\mathcal{S}}^{-1}[\hat{\rho}_{\rm SS}]) = 1$. The fast state is then given by $\hat{\rho}_{\rm f}= \hat{\rho}_{\rm f}(\textrm{p}_{\rm f})=\hat{\mathcal{S}}^{-1}\hat{\rho}_{\rm SS}$ where
\begin{equation}  \label{eq:35}
\textrm{p}_{\rm f} = \frac{\nu-(1-\textrm{p}(\infty))}{\nu-\sigma}.
\end{equation}
If $\textrm{p}_{\rm f} \neq \textrm{p}(\infty)$ and $0 \leq {\rm p}_{\rm f} \leq 1$ then we have a NMQMpE. The evolution of $\hat{\rho}(\tau)$ is given by $\frac{\partial \hat{\rho}(\tau)}{\partial \tau} = \mathcal{L}(\tau)\hat{\rho}(\tau)$ where $\mathcal{L}(\tau) = \sum_{\mu}\hat{F}_{\mu}\hat{G}^{\dag}_{\mu}$ in vectorised form, giving
\begin{equation}  \label{eq:36}
\frac{\partial \textrm{p}(\tau)}{\partial \tau} = \lambda_{2}(\tau)\big(\textrm{p}(\tau)-\textrm{p}_\text{TDFP}(\tau)\big).
\end{equation}
where $\textrm{p}_\text{TDFP}(\tau)$ is the occupation of the time-dependent fixed point of $\hat{\mathcal{L}}(\tau)$. This shows that the evolution is controlled by the instantaneous steady state of $\hat{\mathcal{L}}(\tau)$, acting as an attractor.

\section{Thermofield Transformation and chain mapping} \label{appendix: Thermofield and chain mapping}
The thermofield formalism is a commonly used technique in quantum field theory \cite{takahashi1996thermo,borrelli2021finite,kohn2022quench} and statistical mechanics to relate the properties of a quantum system at finite temperature (impure state) to a doubled system at zero temperature (pure state). For a single fermionic bath mode $\hat{c}$ with Hamiltonian $\hat{H}_{B}=\epsilon\hat{c}^{\dag}\hat{c}$ it's thermal state at inverse temperature $\beta$ and chemical potential $\mu$ is 
\begin{equation}  \label{eq:38}
\rho_{\beta} = \frac{1}{1+e^{-\beta(\epsilon-\mu)}}\big(\ket{0}\bra{0}+e^{-\beta(\epsilon-\mu)}\ket{1}\bra{1}\big),
\end{equation}
where $\ket{1}=\hat{c}^{\dag}\ket{0}$. Introducing an ancilla mode $\hat{a}$ with Hamiltonian $\hat{H}_{A} = \epsilon\hat{a}^{\dag}\hat{a}$, we can express the bath's thermal state as a partial trace of the thermofield double state in the enlarged system as
\begin{equation}  \label{eq:39}
\ket{\Omega_{\beta}}=\sqrt{1-f}\ket{0}_{B}\otimes\ket{0}_{A} + \sqrt{f}\ket{1}_{B}\otimes\ket{1}_{A},
\end{equation}
where $f=(1+e^{\beta(\epsilon-\mu)})^{-1}$ is the fermi factor. Performing a particle hole transformation on the ancilla mode, we have 
\begin{align}  \label{eq:40}
\ket{\Omega_{\beta}} &= \sqrt{1-f}\ket{0}_{B}\otimes\ket{1}_{A}+\sqrt{f}\ket{1}_{B}\otimes\ket{0}_{A}, \nonumber \\
&= \big(\sqrt{1-f}\hat{a}^{\dag}+\sqrt{f}\hat{c}^{\dag}\big)\ket{\text{vac}}. \nonumber
\end{align}
Defining two new fermionic mode operators,
\begin{alignat*}{2}  
\hat{f}^{\dag} = \sqrt{1-f}\hat{a}^{\dag}+\sqrt{f}\hat{c}^{\dag},
\quad&
    \quad &
\hat{e}^{\dag} = \sqrt{f}\hat{a}^{\dag}-\sqrt{1-f}\hat{c}^{\dag},
\end{alignat*}
we see that the thermal state can then be expressed as a single particle product state $\ket{\Omega_{\beta}}=\hat{f}^{\dag}\ket{\text{vac}}$. We now transform our Hamiltonian into this basis. The self energy terms take a simple form,
\begin{equation}  \label{eq:42}
\hat{{H}}_A+\hat{H}_B = \epsilon\big(\hat{a}^{\dag}\hat{a} + \hat{c}^{\dag}\hat{c}\big) 
= \epsilon\big(\hat{f}^{\dag}\hat{f} + \hat{e}^{\dag}\hat{e}\big),
\end{equation}
and if we assume the mode couples to the system via a hybridisation term $\hat{H}_{SB}$ with one system mode $\hat{s}$ we then have
\begin{align}  \label{eq:43}
\hat{H}_{SB} = v\hat{b}^{\dag}\hat{s}+v^{*}\hat{s}^{\dag}\hat{b} = v\sqrt{f}\hat{f}^{\dag}\hat{s} - v\sqrt{1-f}\hat{e}^{\dag}\hat{s}  + v^{*}\sqrt{f}\hat{s}^{\dag}\hat{f} - v^{*}\sqrt{1-f}\hat{s}^{\dag}\hat{e}.
\end{align} 
Moving back to the continuum, $v\to \sqrt{\mathcal{J}_{\alpha}(\omega)}$,$f\to f_{\alpha}(\omega)$, $\hat{c}\to \hat{c}_{\alpha}(\omega)$, we have
\begin{equation}  \label{eq:44}
\hat{H} = \hat{H}_{S}+\sum_{\alpha=L,R}\int_{-D}^{D}\omega\big(\hat{f}_{\alpha}^{\dag}(\omega)\hat{f}_{\alpha}(\omega) + \hat{e}_{\alpha}^{\dag}(\omega)\hat{e}_{\alpha}(\omega)\big)
+ \sqrt{\mathcal{J}_{\alpha f}(\omega)}\big[\hat{f}_{\alpha}^{\dag}(\omega)\hat{s} + \hat{s}^{\dag}\hat{f}_{\alpha}(\omega) \big] 
- \sqrt{\mathcal{J}_{\alpha e}(\omega)}\big[\hat{e}_{\alpha}^{\dag}(\omega)\hat{s} +\hat{s}^{\dag}\hat{e}_{\alpha}(\omega)\big]\mathrm{d}\omega,
\end{equation}
with
\begin{alignat*}{2}  \label{eq:45}
\mathcal{J}_{\alpha f}(\omega) = f_{\alpha}(\omega)\mathcal{J}_{\alpha}(\omega),
\quad&
    \quad &
\mathcal{J}_{\alpha e}(\omega) = \big(1-f_{\alpha}(\omega)\big)\mathcal{J}_{\alpha}(\omega).
\end{alignat*}
This mapping has moved the dependence on temperature from the state into the Hamiltonian, where two separate baths are coupled to the system, one filled and one empty. This separation between the filled and empty modes gives us much greater freedom in optimising our Hamiltonian for matrix product state calculations. The thermal state $\hat{\rho}_{B}$ has become a purified thermofield state $\ket{\Omega_{\beta}}=\prod_{k}\ket{\Omega_{\beta,k}}=\prod_{k}\hat{f}^{\dag}_{k}\ket{\textrm{vac}}$. 

Once the thermofield transformation is applied, we then map each of the continuous baths to finite chains using orthogonal polynomials \cite{Chin_2010,nano11082104}. This is done numerically using the {\tt orthopol} package \cite{2020arXiv200403970M} which implements the following protocol. For simplicity we only consider one bath, but the analysis generalises straightforwardly. We have the following bath Hamiltonian terms 
\begin{equation}  \label{eq:46}
\hat{H}_{SB} = \sum_{c=1,2}\int_{-D}^{D}d\omega \sqrt{J_{c}(\omega)}(\hat{s}^{\dag}\hat{f}_{c}(\omega) + {\rm h.c.}),
\end{equation}
\begin{equation}  \label{eq:47}
\hat{H}_{A}+\hat{H}_{B} = \sum_{c=1,2}\int_{-D}^{D}d\omega\omega\hat{f}_{c}^{\dag}(\omega)\hat{f}_{c}(\omega),
\end{equation}
where we have denoted empty modes as $\hat{e}(\omega)=\hat{f}_{1}(\omega)$ and the filled modes as $\hat{f}(\omega)=\hat{f}_{2}(\omega)$. To carry out the chain mapping, we define new fermionic operators
\begin{equation}  \label{eq:48}
\hat{b}_{c,n} = \int_{-D}^{D}d\omega\sqrt{J_{c}(\omega)}\kappa_{c,n}\pi_{c,n}(\omega)\hat{f}_{c}(\omega),
\end{equation}
with an inverse transformation 
\begin{equation}  \label{eq:49}
\hat{f}_{c}(\omega)=\sum_{n=0}^{\infty}\sqrt{J_{c}(\omega)}\kappa_{c,n}\pi_{c,n}(\omega)\hat{b}_{c,n},
\end{equation}
where $\pi_{c,n}(x)$ is an $n$th monic polynomial with a corresponding normalisation constant $\kappa_{c,n}$ (defined below). We then have $\pi_{c,n}(x) = \sum_{j=0}^{n}c_{nj}x^{j}$, where the monic condition means $c_{nn} = 1$. We define $\pi_{c,n}(x)$ such that they obey the following orthogonality condition 
\begin{equation}  \label{eq:50}
\int_{-D}^{D}d\omega J_{c}(\omega)\pi_{c,m}(\omega)\pi_{c,n}(\omega) = \kappa_{c,n}^{-2}\delta_{n,m},
\end{equation}
which also defines the normalisation constants $\kappa_{c,n}$.
Note that this transformation will leave the state invariant as the new creation (annihilation) operators are linear combinations of creation (annihilation) operators only, so a filled (empty) bath state is mapped to a filled (empty) chain. This is major advantages of the thermofield method for tackling finite temperature. This gives
\begin{align}  \label{eq:51}
\hat{H}_{SB} &= \sum_{c}\sum_{n = 0}^{\infty}\kappa_{c,n}(\hat{s}^{\dag}\hat{b}_{c,n}+ {\rm h.c.}) \int_{-D}^{D}d\omega J_{c}(\omega)\pi_{c,n}(\omega) \nonumber \\
&= \sum_{n = 0}^{\infty}\kappa_{c,n}(\hat{s}^{\dag}\hat{b}_{c,n}+ {\rm h.c.}) \int_{-D}^{D}d\omega J_{c}(\omega)\pi_{c,n}(\omega)\pi_{c,0}(\omega) \nonumber \\
&= \sum_{c}\kappa_{c,0}^{-1}\hat{s}^{\dag}\hat{b}_{c,0}+{\rm h.c.}.
\end{align}
Now consider the bath Hamiltonian
\begin{align}  \label{eq:52}
\hat{H}_{A}+\hat{H}_{B} = \sum_{c=1,2}\sum_{n,m=0}^{\infty}\hat{b}^{\dag}_{c,n}\hat{b}_{c,m}\int_{-D}^{D}d\omega\omega\pi_{c,n}(\omega)\pi_{c,m}(\omega).
\end{align}
To progress, we make use of the following recurrence relation for the monic polynomials.
\begin{equation}  \label{eq:53}
\pi_{c,n+1}(\omega) = (x-\gamma_{c,n})\pi_{c,n}(\omega)- \beta_{c,n}\pi_{c,n-1}(\omega),
\end{equation}
where $\gamma_{c,n}$ and $\beta_{c,n}$ are uniquely determined by the weight function as
\begin{equation}  \label{eq:54}
\gamma_{c,n} = \kappa_{c,n}^{2}\int_{-D}^{D}d\omega\omega J_{c}(\omega)\pi^{2}_{c,n}(\omega),
\end{equation}
and
\begin{equation}  \label{eq:55}
\beta_{c,n} = \kappa_{c,n}\kappa_{c,n+1}\int_{-D}^{D}d\omega J_{c}(\omega)\omega\pi_{c,n}(\omega)\pi_{c,n-1}(\omega).
\end{equation}
Using this, we have
\begin{align}  \label{eq:56}
\hat{H}_{A}+\hat{H}_{B} &= \sum_{c=1,2}\sum_{n,m=0}^{\infty}\kappa_{c,n}^{2}\hat{b}^{\dag}_{c,n}\hat{b}_{c,m}\int_{-D}^{D}d\omega J_{c}(\omega)\pi_{c,m}(\omega)\big[\pi_{c,n+1}(\omega) +\beta_{c,n}\pi_{c,n-1}(\omega) +\gamma_{c,n}\pi_{c,n}(\omega) \big], \nonumber \\
&=\sum_{c=1,2}\sum_{n=0}^{\infty}\bigg(\gamma_{c,n}\hat{b}^{\dag}_{c,n}\hat{b}_{c,n} + \frac{\kappa_{c,n+1}\beta_{c,n+1}}{\kappa_{c,n}}\hat{b}^{\dag}_{c,n}\hat{b}_{c,n+1} + \frac{\kappa_{c,n}}{\kappa_{c,n+1}}\hat{b}^{\dag}_{c,n+1}\hat{b}_{c,n}\bigg),\nonumber \\
&=\sum_{c=1,2}\sum_{n=0}^{\infty}\bigg(\gamma_{c,n}\hat{b}^{\dag}_{c,n}\hat{b}_{c,n} + \sqrt{\beta_{c,n+1}}\hat{b}^{\dag}_{c,n}\hat{b}_{c,n+1} + \sqrt{\beta_{c,n+1}}\hat{b}^{\dag}_{c,n+1}\hat{b}_{c,n}\bigg).
\end{align}

Combining this with the rest of the Hamiltonian and choosing a finite cutoff $N_{b}$ for both baths, we obtain 
\begin{align}  \label{eq:57}
\hat{H} &= \hat{H}_{S}+\sum_{\tilde{\alpha}}\kappa_{\tilde{\alpha},0}^{-1}(\hat{s}^{\dag}\hat{b}_{\tilde{\alpha},0}+\hat{b}^{\dag}_{\tilde{\alpha},0}\hat{s}) \nonumber\\
&\qquad\qquad+\sum_{\tilde{\alpha}}\sum_{n=0}^{N_{n}}\bigg(\gamma_{\tilde{\alpha},n}\hat{b}^{\dag}_{\tilde{\alpha},n}\hat{b}_{\tilde{\alpha},n} + \sqrt{\beta_{\tilde{\alpha},n+1}}\hat{b}^{\dag}_{\tilde{\alpha},n}\hat{b}_{\tilde{\alpha},n+1} + \sqrt{\beta_{\tilde{\alpha},n+1}}\hat{b}^{\dag}_{\tilde{\alpha},n+1}\hat{b}_{\tilde{\alpha},n}\bigg),
\end{align}
where $\tilde{\alpha}$ runs over all combinations of $\alpha,c$. 

\section{Correlation matrix propagation} \label{appendix: corr propagation}
Here we prove Eq.~(\ref{eq:20}). Defining $\mathcal{U}(\tau) = e^{-i\hat{H}\tau}$, we have
\begin{align} \label{eq:58}
\boldsymbol{C}_{ij}(\tau) &= \bra{\psi(\tau)}\hat{c}^{\dag}_{j}\hat{c}_{i}\ket{\psi(\tau)} \nonumber = \bra{\psi(0)}\mathcal{U}^{\dag}(\tau)\hat{c}^{\dag}_{j}\hat{c}_{i}\mathcal{U}(\tau)\ket{\psi(0)}.
\end{align}
To evaluate this, first consider how $\hat{H}$ acts on $\hat{c}^{\dag}_{j}$,
\begin{align} 
\hat{H}\hat{c}^{\dag}_{j} &= \sum_{kl}\boldsymbol{H}_{kl}\hat{c}^{\dag}_{k}\hat{c}_{l}\hat{c}^{\dag}_{j} = \sum_{kl}\boldsymbol{H}_{kl}\hat{c}^{\dag}_{k}(\delta_{lj} - \hat{c}^{\dag}_{j}\hat{c}_{l}) = \sum_{k}\boldsymbol{H}_{kj}\hat{c}^{\dag}_{k}.
\end{align}
We can now evaluate $\mathcal{U}^{\dag}\hat{c}^{\dag}_{j}$ as
\begin{align}  \label{eq:60}
\mathcal{U}^{\dag}\hat{c}^{\dag}_{j} &= e^{iHt}\hat{c}^{\dag}_{j} = e^{i\sum_{k}\boldsymbol{H}_{kj}t}\hat{c}^{\dag}_{k} = \sum_{k}\boldsymbol{U}^{\dag}_{kj}\hat{c}^{\dag}_{k},
\end{align}
Substituting this into Eq.~(\ref{eq:58}) gives the result
\begin{align}  \label{eq:61}
\boldsymbol{C}_{ij}(t) &= \bra{\psi(0)}\sum_{kl} \boldsymbol{U}^\dagger_{kj}\hat{c}^\dagger_{k}(\boldsymbol{U}^\dagger_{li}\hat{c}^\dagger_{l})^\dagger \ket{\psi(0)} = \sum_{kl} \boldsymbol{U}^\dagger_{kj}\boldsymbol{U}_{il}\bra{\psi(0)}\hat{c}^{\dagger}_{k}\hat{c}_{l}\ket{\psi(0)} = \sum_{kl} \boldsymbol{U}^\dagger_{kj}\boldsymbol{U}_{il} \boldsymbol{C}_{lk}(0).
\end{align}
Thus we have
\begin{equation}  \label{eq:62}
\boldsymbol{C}(t) = \boldsymbol{U}\boldsymbol{C}(0)\boldsymbol{U}^{\dag}.
\end{equation}

\end{document}